\documentclass[authoryear,preprint,review,12pt,fleqn]{elsarticle}
\usepackage[utf8]{inputenc}
\usepackage[a4paper, total={6in, 10in}]{geometry}
\usepackage{xcolor}
\usepackage{amsmath}
\usepackage{mathtools}
\usepackage{amssymb}
\usepackage{caption}
\usepackage{textcomp}
\usepackage{booktabs}
\usepackage{subcaption}
\usepackage{graphicx}
\usepackage{graphics}
\usepackage{soul}
\usepackage{multicol}
\usepackage{csquotes}
\usepackage{natbib}
\usepackage{float}
\usepackage{adjustbox}
\usepackage{tikz}

\usepackage{lineno}

\journal{xx}

\begin{document}

\begin{frontmatter}



\title{Thermal Acoustical and Mechanical Characterization of Rock under in-situ conditions}


\author[1]{Nima Haghighat\corref{cor1}}
\ead{nima.haghighat@ifg.uni-kiel.de}
\author[1]{Hem B. Motra}
\author[1]{Amir S. Sattari}
\author[1]{Frank Wuttke}

\cortext[cor1]{Corresponding author}
\affiliation[1]{organization={Geomechanics \& Geotechnics Department, Kiel University},
            addressline={24118, Kiel}, 
            country={Germany}}

\begin{abstract}
Data on the coupled influence of pressure and temperature on rock thermal conductivity remain limited. This study adapts a multi-anvil cubic press to enable steady-state thermal conductivity measurements under controlled in-situ conditions. The setup concurrently records P- and S-wave velocities along three orthogonal directions, providing unique real-time constraints on micro-structural evolution. Beyond the expected temperature dependence, preliminary tests on two sandstones reveal a twofold pressure effect: increasing confining pressure enhances thermal conductivity through microcrack closure and better grain contacts, while simultaneously reducing the rate at which conductivity decreases with temperature. Initial results align closely with published datasets, supporting the reliability of the developed methodology. Quantitatively, the thermal conductivity of the studied sandstones was found to decrease by an average of approximately 5.9\% per $50~^{\circ}\text{C}$ at the lowest confinement (12~MPa), compared with about 4.2\% per $50~^{\circ}\text{C}$ at the highest confinement (100~MPa). In addition, an average increase of roughly 1.4\% per 10~MPa
of applied confining pressure was observed for both samples.

\end{abstract}


\end{frontmatter}


\section{Introduction}

Earth’s internal heat is transmitted to the surface primarily through radiation, advection, and conduction, with conduction being the dominant mechanism within the lithosphere \citep{clauser1995thermal}. Reliable in-situ thermal conductivity data are therefore indispensable across many geophysical and engineering applications, including geothermal energy assessment \citep{barbier2002geothermal,balling2013deep,velez2018geothermal}, underground thermal energy storage \citep{sanner2003current,bonte2011underground,sadeghi2022energy}, hydrocarbon reservoir management \citep{abid2014thermophysical,xu2016review,askarova2020thermal}, and the safe disposal of radioactive waste \citep{tsang2012coupled,xu2019temperature}. Quantifying the spatial distribution of thermal conductivity in the lithosphere necessitates knowledge of how crustal rocks conduct heat under temperature and pressure conditions that vary with depth, as determined from laboratory measurements~\citep{seipold1995variation}. Most studies of rocks’ thermal properties under crustal conditions have treated pressure and temperature effects separately, highlighting temperature as a key factor controlling thermal conductivity. Although the lattice thermal conductivity of a structurally perfect, isotropic single crystals is theoretically expected to vary inversely with temperature, most rocks, being composed of complex, disordered crystalline structures with diverse mineral compositions, do not exhibit a simple temperature dependence. \cite{birch1940thermal,birch1940thermal2} investigated the thermal conductivity of various igneous, sedimentary, and metamorphic rocks over a temperature range of $0~^{\circ}\text{C}$ to $500~^{\circ}\text{C}$. Their results showed that the thermal conductivity of most of the studied rock types decreases with increasing temperature. Building upon the collection of available data in the literature (e.g.,~\cite{sugawara1962experimental, beck1978lattice, mongelli1982thermal}), \cite{lee1998evaluation} demonstrated that rocks and minerals with a thermal conductivity greater than 2 $\text{W}~\text{m}^{-1}~\text{K}^{-1}$ at room temperature typically experience a decline in conductivity as temperature increases, while those with a lower conductivity tend to see an increase. The temperature dependence of rock thermal conductivity has consistently drawn research interest~\citep{vosteen2003influence,mottaghy2008temperature, miao2014temperature,wen2015temperature,miao2018temperature}. More recently, \cite{chen2021effect} explored the thermal conductivity of various rock types at elevated temperatures using the Hot Disk method. Their results confirmed that the trend in thermal conductivity depends heavily on the material’s room temperature conductivity, identifying three distinct regions: rocks with conductivity above 4.5~$\text{W}~\text{m}^{-1}~\text{K}^{-1}$ at $25~^{\circ}\text{C}$ showed a sharp decline in conductivity as the temperature rose; those with conductivity between 2.5$\text{W}~\text{m}^{-1}~\text{K}^{-1}$  and 3.5~$\text{W}~\text{m}^{-1}~\text{K}^{-1}$  decreased linearly; and samples with a conductivity below 2.5$\text{W}~\text{m}^{-1}~\text{K}^{-1}$  initially increased conductivity up to $150~^{\circ}~\text{C}$ before declining at higher temperatures.\\
The effect of pressure on conductivity, on the other hand, is largely dictated by mineral composition, porosity, and density, with a general trend of increasing conductivity as pressure rises. \cite{horai1989effect} examined thermal conductivity changes under different pressure regimes, noting that the most pronounced increase occurs as pressure rises up to 2~$\text{kbar}$. Beyond 2~$\text{kbar}$, thermal conductivity tends to increase more linearly for most rock types. The general pattern of a sharp initial increase followed by a gradual linear rise in the thermal conductivity of rocks has been confirmed by other studies. \cite{abdulagatov2006effect} attributed the increase in rock conductivity under pressure to the closure of fractures and microcracks upon bringing rocks to the surface, and the enhanced mechanical contact between grains, which reduces internal thermal resistance. According to \cite{adams1923compressibility}, very narrow crack-like openings along grain boundaries form the majority fraction of the porosity in low-porosity rocks which are expected to close under compressive stress. Therefore, even for low-porosity rocks, the effect of pressure on the thermal conductivity should not be neglected~\citep{walsh1966effect}. In this regard, per 10~$\text{MPa}$, a 1.3~\% rate of increase in thermal conductivity of low porosity rocks has been reported~\citep{lebedev1988temperature}.\\
As stated earlier, the majority of studies dedicated to the evaluation of the thermal conductivity of rocks focus either on the sole effect of temperature under atmospheric pressure or on the effect of pressure at a constant temperature, rarely addressing the combined influence of both factors. Early examples of studies dedicated to the investigation of rocks' thermal properties under the coupled effect of pressure and temperature at crustal conditions can be found in \cite{buntebarth1987laboratory}, and \cite{seipold1992depth}. In these studies, the effects of increasing pressure and temperature on the thermal characteristics of the rocks are initially examined independently. Following this, the combined influence of pressure and temperature is explored by introducing the concept of depth, where the two factors are superposed to reflect different crustal conditions. It is widely recognized that, at elevated temperatures and ambient pressure, differences in thermal expansion between adjacent mineral grains can lead to the formation of thermally induced cracks. Therefore, in the absence of increasing pressure, which would otherwise close thermally induced cracks, one might expect that the resulting temperature-dependent measurements of thermal conductivity are underestimated. Recognizing these limitations, recent research has increasingly focused on the simultaneous and coupled effects of increasing temperature and pressure on rock thermal conductivity. By means of a steady-state guarded parallel plate apparatus, \cite{abdulagatova2009effect,abdulagatova2010effect} and \cite{emirov2021studies} studied the temperature-baric dependence of sandstone thermal conductivity at elevated conditions. The common highlight of these studies is the fact that the heat transfer in rock is sensitive to lattice defects and grain boundaries. As a result, pressure not only influences thermal conductivity directly but
also alters its temperature dependence by affecting the internal structure and contact between mineral grains. This further emphasizes the need to examine the combined effects of temperature and pressure, particularly in disordered, polycrystalline rock materials. 
Despite its importance, the combined influence of pressure and temperature on rock thermal conductivity has been investigated in only a limited number of studies. Therefore, the objective of the present work is to establish a methodology for evaluating the coupled temperature–pressure dependence of thermal conductivity in rocks. To achieve this, the multianvil cubic press apparatus at the Geomechanics and Geotechnics Department of Kiel University was adapted to allow steady-state thermal conductivity measurements under controlled pressure and temperature conditions.

Furthermore, acknowledging the well-documented correlation between thermal conductivity and elastic wave velocities in many rock types~\citep{ozkahraman2004determination,el2019thermal}, the experimental setup was designed to enable unique simultaneous measurements of P- and S-wave velocities in three principal directions in conjunction with thermal conductivity measurements. Variations in P-wave velocities primarily reflect changes in bulk density and compressional stiffness associated with porosity evolution, whereas measurements of polarized S-wave velocities along different orientations can be indicative of stress-aligned microcrack closure or development during combined heating and compression~\citep{kern1993p}. This dual-measurement approach offers a unique means of tracking microstructural changes in real time as temperature and pressure vary, thereby improving the interpretation of the observed thermal response. To ensure validation and calibration of this novel methodology and to establish a benchmark against standardized thermal conductivity measurements, an additional experimental method was designed and configured based on a customized oedometer cell concept, closely resembling conventional techniques reported in the literature.

The structure of the present study is organized as follows: first, the two experimental methodologies employed for the evaluation of rock thermal conductivity under in-situ conditions are described. Subsequently, the studied sample and the results of the characterization techniques applied to these samples are presented. Finally, the thermal response of the sample is analyzed, followed by a discussion of the results and a comparison with available analytical and experimental data reported in the literature.

\section{Methodology}

As outlined in the introduction, to validate and benchmark the primary experimental approach, that is, direct thermal conductivity measurement under true-triaxial stress using the cubic press, a complementary experimental setup was developed. This system is based on the concept of one-dimensional temperature profiles with guard heaters combined with uniaxial mechanical loading, closely resembling conventional thermal conductivity measurement techniques.  To realize this goal and provide a versatile and adaptable experimental platform, a customized cylindrical oedometer cell (COC) was successfully designed and constructed. This apparatus is capable of operating at temperatures up to $200~^{\circ}\text{C}$ and enables independent thermal control at the top, bottom, and lateral boundaries of the sample. The primary motivation behind this design is to facilitate thermal conductivity measurements using the comparative steady-state method, which requires a stable and controllable one-dimensional heat flow through the sample while minimizing lateral heat losses. Inspired by \cite{sass2012coupled}, the design enables the determination of the sample's thermal conductivity by comparing the temperature drop across the sample with that across a reference material. In this study, a zirconia-based ceramic ($\text{ZrO}_2$), specifically TZP (DIN designation), supplied by BCE Special Ceramics, was used as the reference material due to its well-characterized and stable thermal properties ($\text{k} = 2~\text{W}~\text{m}^{-1}~\text{K}^{-1}$).

\begin{figure}[h!]
  \centering
  \begin{subfigure}[b]{0.48\textwidth}
    \includegraphics[trim=0cm 0.5cm 0cm 0.1cm, clip=true, width=\textwidth]{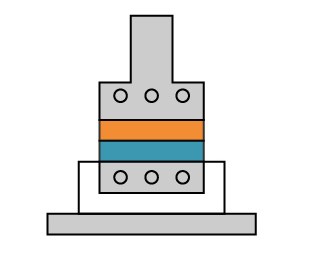}
    \subcaption{}
  \end{subfigure}
  \hfill
  \begin{subfigure}[b]{0.35\textwidth}
    \includegraphics[width=\textwidth]{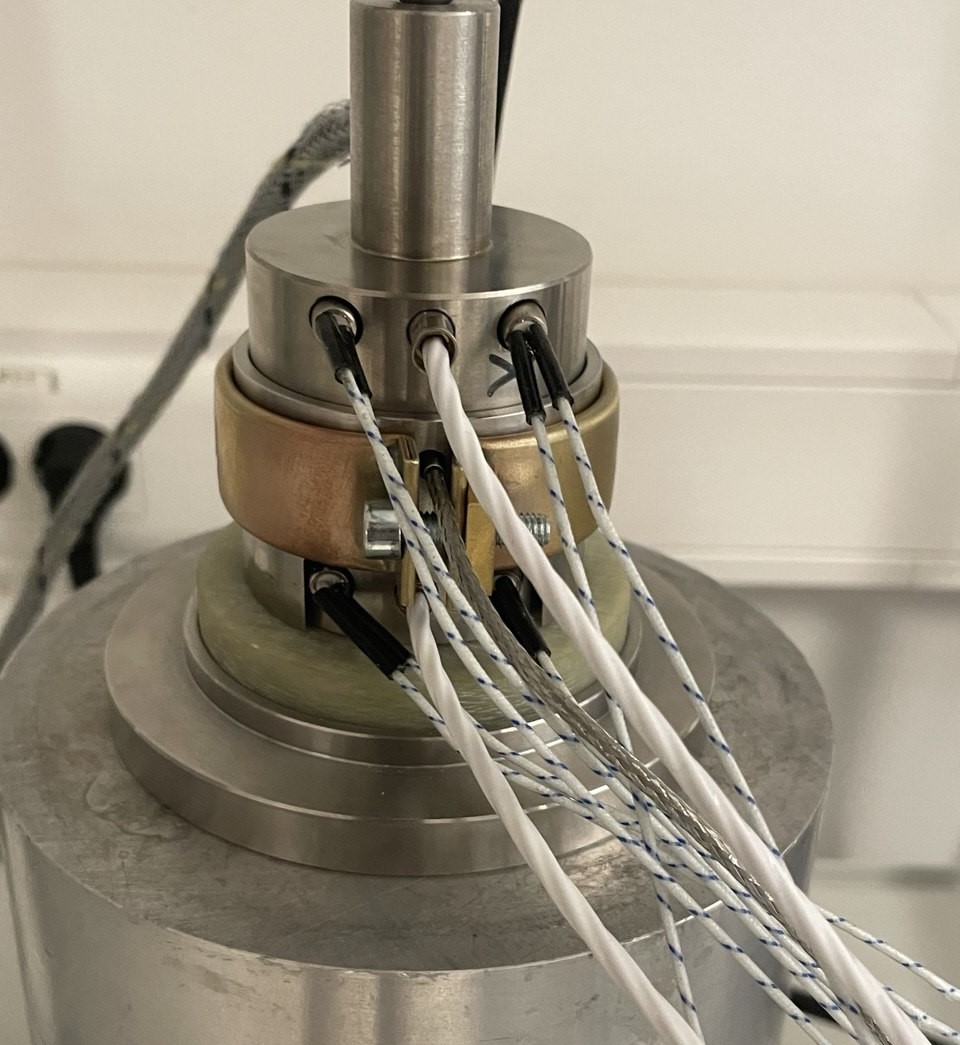}
    \subcaption{}
  \end{subfigure}
  \caption{(a) 2D schematic design of the COC, including cross-sectional details of the top and bottom heating caps; (b) detailed view of the assembled COC setup in its physical configuration.}
      \label{fig:COC_SETUP}
\end{figure}

The schematic representation and physical implementation of the COC are shown in Fig.~\ref{fig:COC_SETUP}. The general configuration consists of a vertically stacked assembly comprising the studied sample and reference material, sandwiched between top and bottom heating caps. Both the top and bottom caps are equipped with two heating coils and a PT100 temperature sensor positioned at the center, enabling precise temperature regulation via PID controllers. The temperature of the surrounding ring heater, which encloses the stack, is also independently regulated to minimize radial heat losses. In addition, the bottom heating cap is thermally isolated from the base of the cell using a low-conductivity insulating material. During the measurements, to further minimize external heat losses, the entire heating system is enclosed within a two-layer insulation system. The inner layer consists of a fibrous ceramic insulation, which is surrounded by an outer cast made of a low-conductivity material. This dual-layer insulation design ensures thermal stability and enhances the accuracy of the thermal conductivity measurements by reducing uncontrolled heat exchange with the environment.

All cylindrical components, including the sample, reference materials, and heating caps have a diameter of 5 cm. The height of the sample and the reference material is 1 cm. To perform the measurements, a constant temperature gradient is applied across the stack by independently controlling the temperatures of the top and bottom heating caps. The side heater is maintained at the average temperature of the heating caps to ensure a linear heat profile. The temperature response is then measured at the center of the reference material using a 1/3 DIN PT100 temperature sensor, with data logged via an HBM MGCplus system. The temperatures of the heating caps are monitored and recorded through the Modbus communication protocol, using a Python script to query and log the temperature sensors. The obtained thermal conductivity measurements using the COC setup provides a reference for evaluating the accuracy and reliability of the cube press methodology described in the following section.

\begin{figure}[t!]
    \centering
    \hspace{20pt}
    \begin{minipage}{0.45\textwidth}
        \centering
        \includegraphics[width=\linewidth]{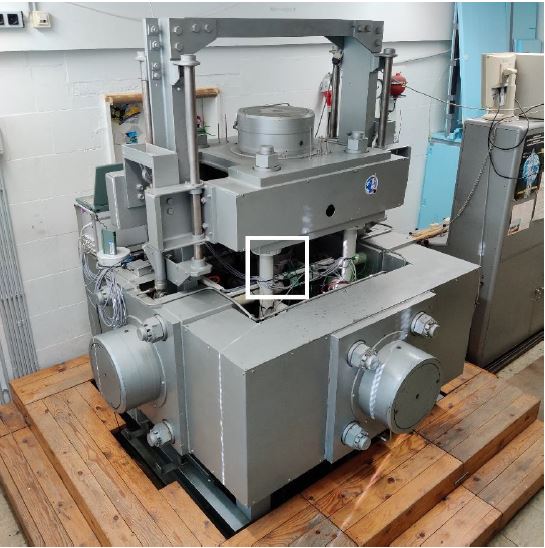}
        \vspace{10pt}
        \subcaption{\label{fig:WP_Setup}}
    \end{minipage}
    \hspace{10pt}
    \begin{minipage}{0.45\textwidth}
        \centering
        \vspace{50pt}
        \includegraphics[width=\linewidth]{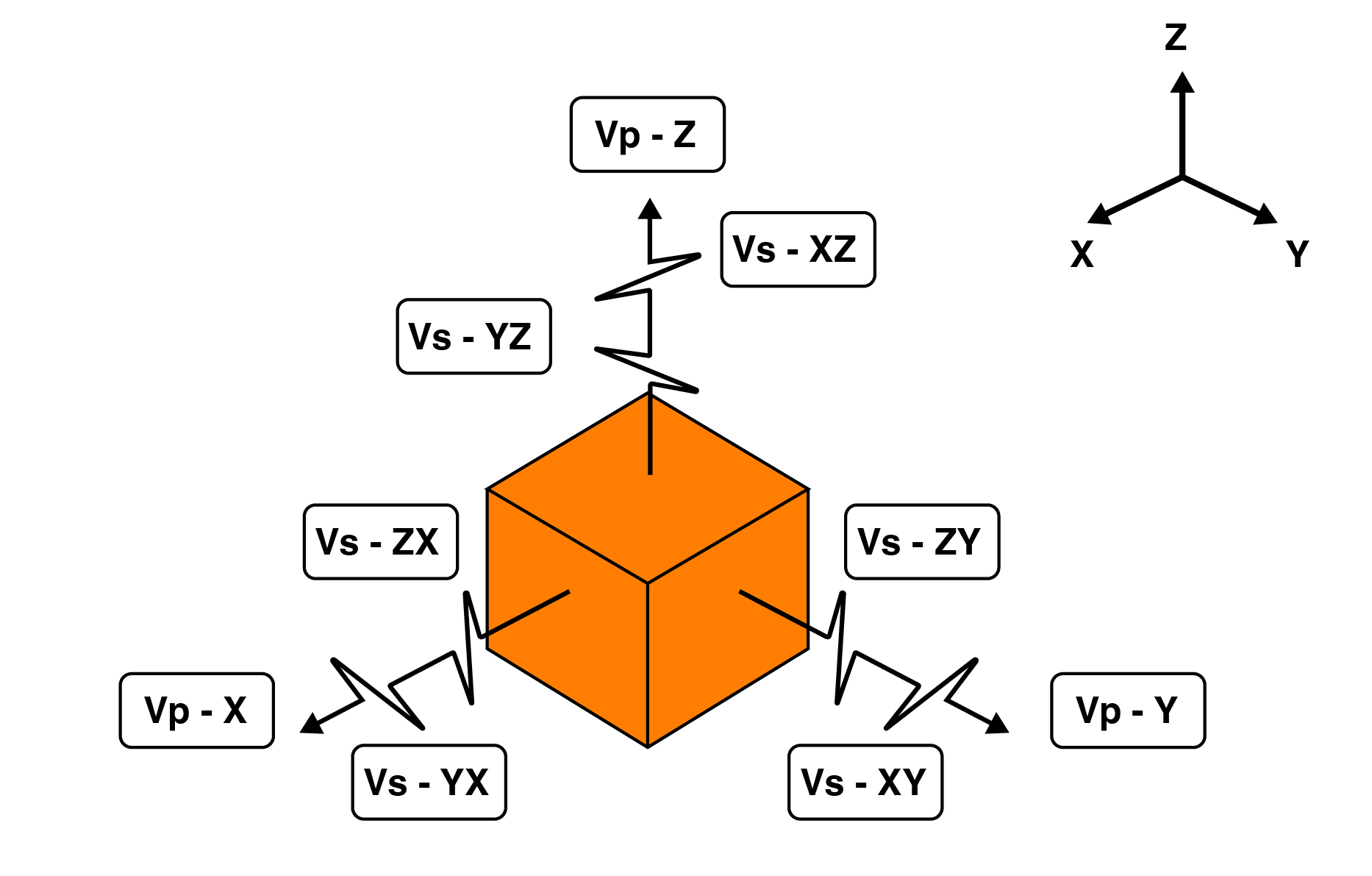}
        \vspace{40pt} 
        \subcaption{\label{fig:WP_WaveSchematics}}
    \end{minipage}
    \caption{(a) The cubic multi-anvil pressure apparatus. (b) Schematic description of measured P- and S-wave velocities.}
    \label{fig:WP_overall}
\end{figure}

For simultaneous measurements of rock thermal conductivity and three-dimensional elastic wave velocities, the multi-anvil cubic press in the Geomechanics and Geotechnics department of Kiel University was employed. The device, as shown in Fig.~\ref{fig:WP_Setup}, is capable of
applying pressures up to 600 MPa and temperatures up to $600~^\circ\text{C}$ in three principal directions. Further details on the design and operation of the cubic press can be found in \cite{kern1990fabric, kern1993p}. In brief, the apparatus applies quasi-hydrostatic stress by means of six opposed pyramidal pistons acting on a cubic specimen, with an edge length of 43~mm. Heating is achieved by furnaces integrated into the pistons, ensuring a homogeneous temperature distribution within the specimen. Temperature is monitored using thermocouples positioned within 1 mm of the sample surface, while ultrasonic P- and S-wave velocities are measured using the pulse-transmission technique. Ceramic transducers with a 2 MHz resonant frequency are mounted on the cooler side of the pistons. Wave travel times are obtained by correcting the measured signals for the piston paths, and velocities are adjusted for changes in sample dimensions with pressure and temperature. The total deformation in a given direction is calculated by summing the average displacements from the two sensors on each opposing piston cap. As shown in Fig.~\ref{fig:WP_WaveSchematics}, to assess the shear-wave birefringence, two orthogonally polarized shear-wave transducers were employed in different orientations. The overall timing precision is ±5~ns, corresponding to an accuracy better than 0.5\%.

For thermal conductivity determination, a comparative steady-state approach was employed, similar in principle to the previously described COC setup. In this method, the temperature drop across the specimen is compared to that of a reference material with a well-characterized thermal conductivity. A synthesized zirconia plate served as the reference and was placed directly above the sample within the assembly. The constant thermal gradient was established by regulating the temperatures of the top and bottom pistons, while the surrounding pistons functioned as guard heaters to minimize lateral heat losses. The temperature drop within the reference material was monitored at its center using a resistance temperature sensor.

\subsection{Calibration and Reference Testing}
This section describes the procedures implemented to evaluate the robustness and reliability of the measurement systems, along with the calibration strategies adopted prior to experimental testing. These steps, outlined in detail below, were critical to ensuring the accuracy and consistency of measurements under high-pressure and high-temperature conditions. For the COC-based setup, validation focused exclusively on thermal conductivity measurements, whereas for the cubic press, both elastic wave velocity and thermal conductivity measurements were calibrated and verified through multiple tests.

\subsubsection{Customized Oedometer Cell}

To assess the accuracy of the COC measurement system, a series of calibration tests was conducted using a quartz glass sample with well-established thermal conductivity values. Quartz glass is a common standard in thermal conductivity validation due to its homogeneity and predictable thermal behavior. The test procedure starts with the placement of the stack, consisting of the reference material and the studied sample, inside the cell, which is then mounted within the loading framework. Next, all necessary connections for heating and temperature logging are established. The sample and reference material are loaded under a constant pressure before the setup is enclosed with dual-layer insulation. After reaching mechanical equilibrium under applied load, in this case 100~$\text{kPa}$, a temperature gradient is applied by maintaining a $10~^\circ\text{C}$ difference between the top and bottom heating caps. Thermal conductivity measurements are then conducted over an average temperature range of $50~^\circ\text{C}$ to $150~^\circ\text{C}$ in $25~^\circ \text{C}$ increments. At each step, the system is allowed to reach thermal equilibrium, typically taking about one hour, before advancing to the next temperature level.

According to Fourier’s law, and assuming one-dimensional heat flow while neglecting radiative heat transfer terms, the rate of heat flow per unit area, $q$, is proportional to the negative temperature gradient and can be expressed in terms of thermal conductivity, \emph{i.e.}, $k$, as:
\begin{equation}
    q = -k\nabla T= -k\frac{\Delta T}{\Delta S}
\end{equation}
where $\Delta S$ is the distance between temperature gradient points. Upon reaching steady-state heat transfer conditions, it can be assumed that the heat flux between the top cap and the bottom cap is equal to that between the measuring point within the reference material and the bottom cap. So one can write:
\begin{equation}
    -\frac{T_2-T_3}{S_{23}/k_v} = -\frac{T_1 - T_3}{S_v/k_v + S_p/k_s}
    \label{eq:FlowOneDimensional}
\end{equation} where $T_1$, $T_3$, and $T_2$ are the temperatures of the top cap, bottom cap, and the measuring point within the reference material, respectively. $S_v$, $S_p$, and $S_{23}$ are the heights of reference material, sample, and the distance between measuring point $T_2$ and the bottom cap. $k_v$ is the thermal conductivity of the reference material. Solving for Eq.~\ref{eq:FlowOneDimensional} would result in thermal conductivity of the sample, $k_s$, as:
\begin{equation}
    k_s = \frac{k_v S_p}{\frac{T_1-T_3}{T_2-T_3}S_{23}-S_v}
    \label{eq:TC1D}
\end{equation}

\begin{figure}[t!]
    \centering
    \includegraphics[width=0.7\linewidth]{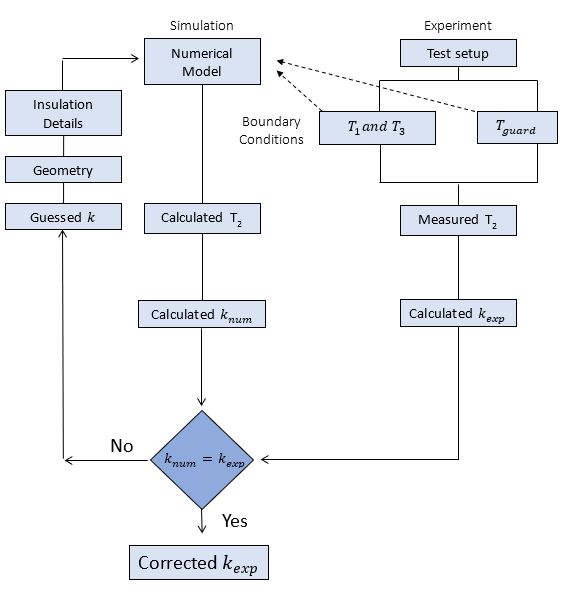}
    \caption{Flowchart illustrating the correction procedure for determining true thermal conductivity using iterative numerical modeling; $k_{num}$ and $k_{exp}$ are the numerically and experimentally derived thermal conductivities based on Eq.~\ref{eq:TC1D}.}
    \label{fig:CorrectionScheme}
\end{figure}

\noindent Eq.~\ref{eq:TC1D} is expected to yield accurate results under ideal conditions, e.g., zero thermal conductivity of the insulation material. However, such conditions are seldom achieved in practical applications. The total uncertainty in the measured thermal conductivity stems from both measurement-related errors, such as sensor accuracy and temperature logging precision, and systematic errors, including heat losses, imperfect thermal boundary conditions, and deviations from the idealized one-dimensional heat transfer assumption. To minimize measurement-related errors, several precautions were taken during the experiments. As previously mentioned, a high-precision data acquisition system (HBM MGCplus) was employed to ensure accurate and stable temperature logging. Additionally, thermal paste was applied around the temperature sensors to improve thermal contact and reduce interface resistance. To address systematic effects, on the other hand, and enhance the reliability of the results, a finite element analysis-based correction procedure was implemented following the methodology described in \cite{xing2014correction}. In this way, by performing numerical simulations that replicate the experimental boundary conditions, the true thermal conductivity is obtained through an optimization process that aligns the simulation results with experimental data. The involved steps of the optimization procedure are outlined in Fig.~\ref{fig:CorrectionScheme}. 

\begin{figure}[htbp]
    \centering
    \includegraphics[width=0.65\linewidth]{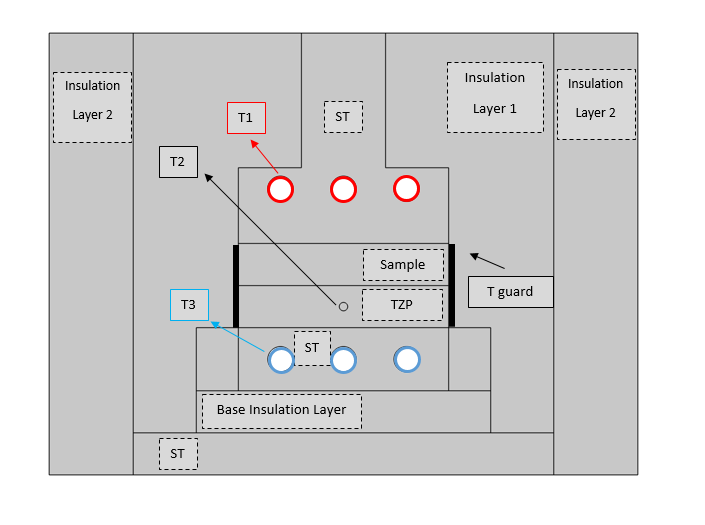}
    \caption{The numerical replica representing the COC thermal conductivity measurement device; The sample is positioned between two steel (ST) caps, while a zirconia-based ceramic plate (TZP) is placed below the sample and serves as the reference material for the comparative steady-state method. }
    \label{fig:COC_Comsol_Setup}
\end{figure}

To apply the correction scheme and quantify systematic errors in thermal conductivity measurements, a numerical model was developed using the finite element software $\text{COMSOL Multiphysics}^{\textregistered}$. As shown in Fig.~\ref{fig:COC_Comsol_Setup}, the model represent the experimental geometrical details and thermal boundary conditions of the COC setup. The top and bottom caps, as well as the surrounding surfaces of the sample and reference material, are defined with fixed temperature boundaries, corresponding to the set temperatures during experiments. The required material properties, including temperature-dependent thermal conductivities, were assigned based on available literature and manufacturer data. As mentioned earlier, the inner insulation layer consists of insulation wool, with a thermal conductivity of 0.04~$\text{W}~\text{m}^{-1}~\text{K}^{-1}$ at $50~^{\circ}\text{C}$ and 0.085~$\text{W}~\text{m}^{-1}~\text{K}^{-1}$ at $300~^{\circ}\text{C}$. The outer layer is made of a low-conductivity casing material with a thermal conductivity of approximately 0.1~$\text{W}~\text{m}^{-1}~\text{K}^{-1}$. For both the inner and outer insulation layers, average thermal conductivity values were assigned in the numerical model. To accurately capture the measurement conditions, the temperature sensors were represented as small cylindrical volumes with a diameter of 3~$\text{mm}$ and a thermal conductivity of 4.5~$\text{W}~\text{m}^{-1}~\text{K}^{-1}$~\citep{elkholy2022accurate}. The temperature drop across the sample was extracted from the simulation by computing the surface-averaged temperature at the sensor . This simulated response was then iteratively compared to the experimentally measured values. The thermal conductivity of the sample was adjusted within the model until the simulated and experimental temperature drops converged.

\subsubsection{Cube press}

\begin{figure}[htbp]
    \centering
    \begin{subfigure}[b]{0.45\textwidth}
        \centering
        \includegraphics[width=\textwidth]{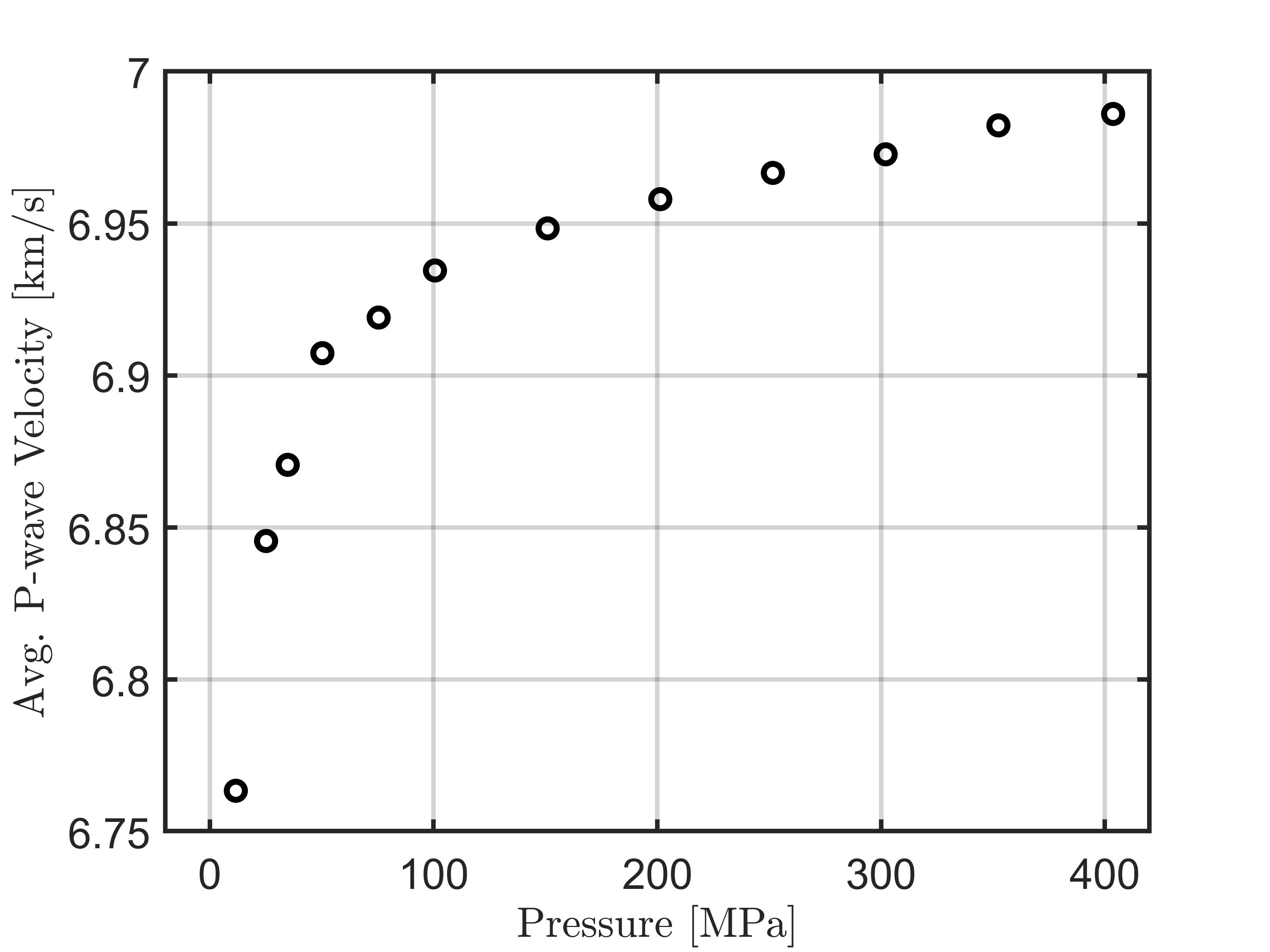}
        \subcaption{}
    \end{subfigure}
    \hfill
    \begin{subfigure}[b]{0.45\textwidth}
        \centering
        \includegraphics[width=\textwidth]{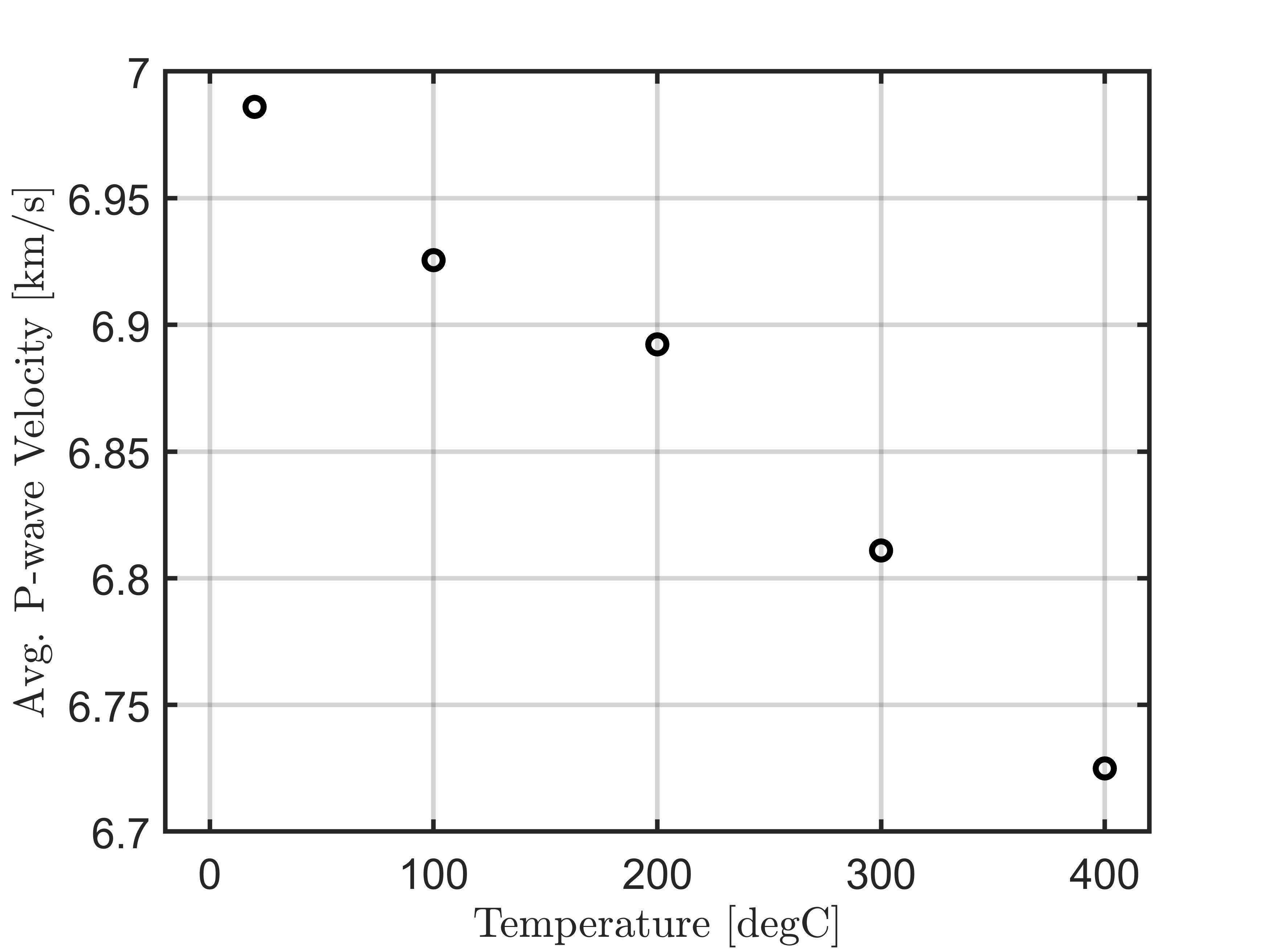}
        \subcaption{}
    \end{subfigure}

    \vskip\baselineskip

    \begin{subfigure}[b]{0.45\textwidth}
        \centering
        \includegraphics[width=\textwidth]{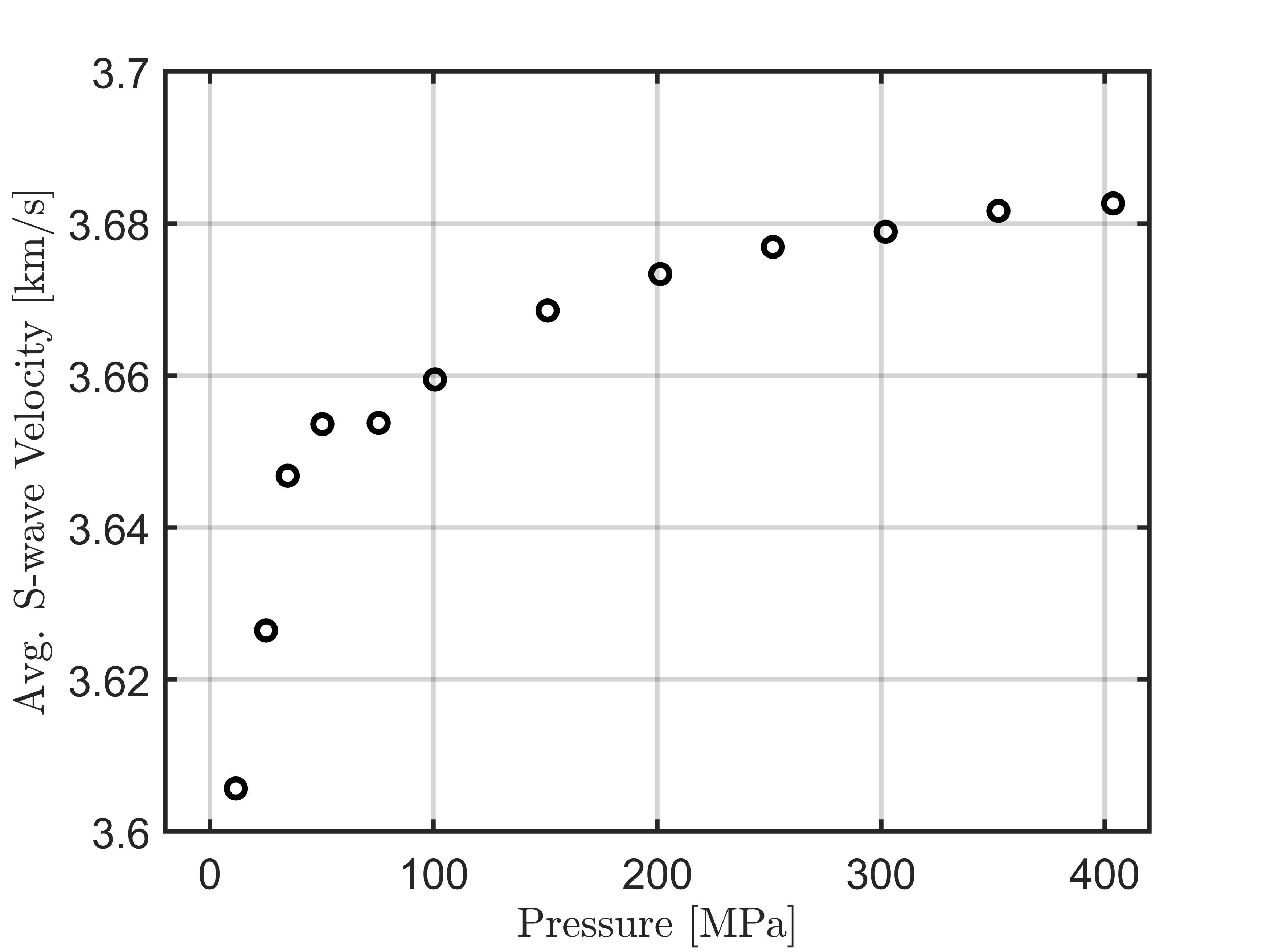}
        \subcaption{}
    \end{subfigure}
    \hfill
    \begin{subfigure}[b]{0.45\textwidth}
        \centering
        \includegraphics[width=\textwidth]{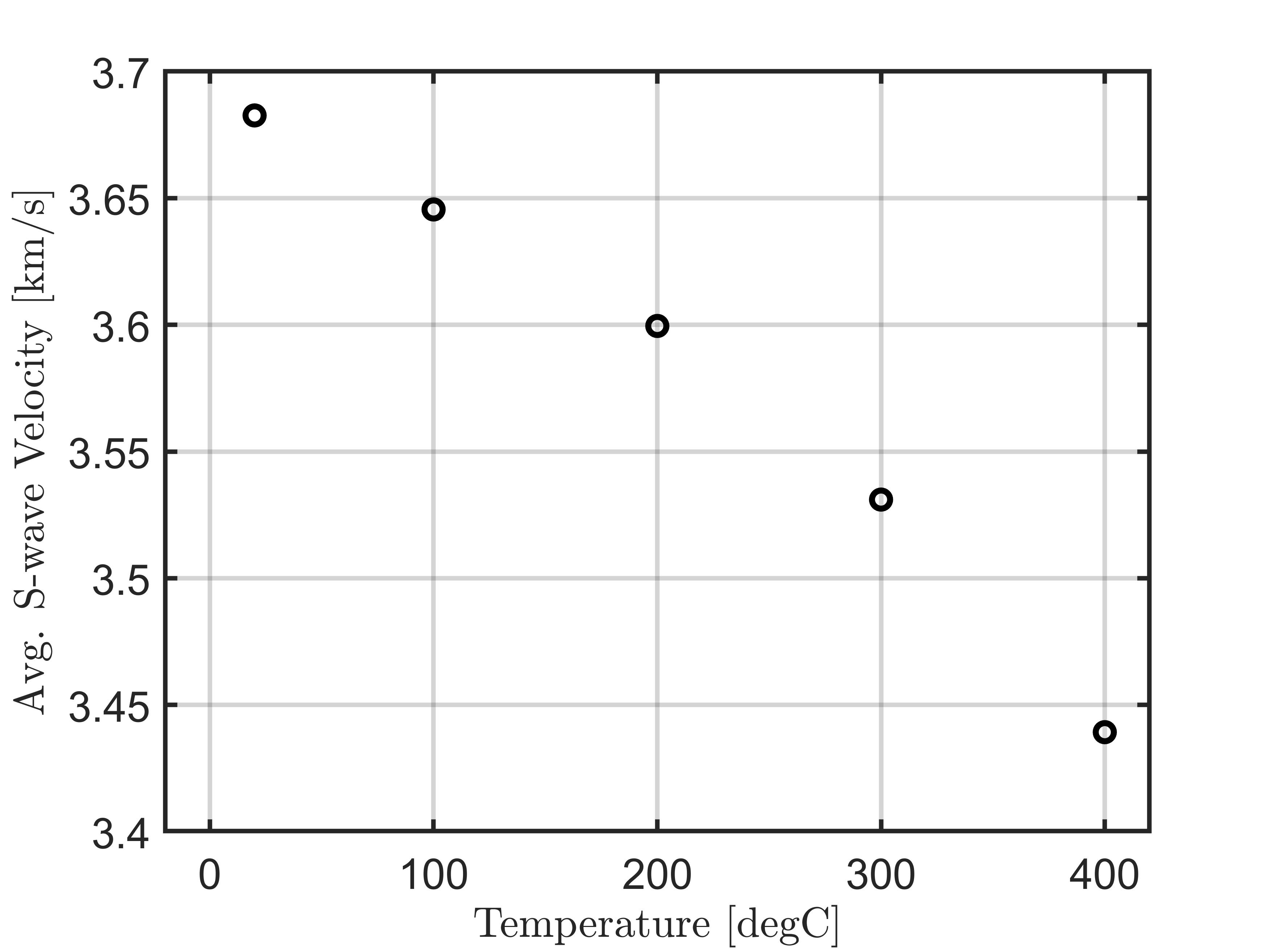}
        \subcaption{}
    \end{subfigure}
    
    \caption{(a) Average P-wave velocities of the reference material (TZP) under varying pressure and (b) temperature conditions. (c) Average S-wave velocities of the reference material (TZP) under varying pressure and (d) temperature conditions.\label{fig:ZR700-PresTempEffect}}
\end{figure}

To evaluate the robustness of the wave velocity measurements, a series of experiments was conducted in which P-wave and S-wave signals were recorded for the employed reference material (TZP) across a range of pressures and temperatures. In addition to serving validation purposes, these measurements provide essential reference values under varying stress and temperature conditions. These values are further used in the interpretation of wave velocity data obtained simultaneously during the thermal conductivity experiments on the studied samples. The P-wave and S-wave velocities were evaluated in three main orthogonal directions under different stress and temperature regimes. Here, the average measured P-wave and S-wave velocities under quasi-hydrostatic pressures ranging from 12 to 400~MPa at constant room temperature, as well as under increasing temperatures from room temperature up to $400~^{\circ}\text{C}$ at a constant applied pressure of 400~MPa, are illustrated in Fig.~\ref{fig:ZR700-PresTempEffect}. As can be seen, pressure contributes positively to the average P- and S-wave velocities, with velocities increasing as pressure rises. In contrast, increasing temperature leads to a decrease in both P-wave and S-wave velocities.

The anisotropy ratio, $AR$, could be evaluated using

\begin{equation}
AR = \frac{V_{p_{\text{max}}} - V_{p_{\text{min}}}}{(V_{p_x} + V_{p_y} + V_{p_z})/3}
\label{eq:anisotropy}
\end{equation}

where $V_{p_{\text{max}}}$ and $V_{p_{\text{min}}}$ denote the maximum and minimum P-wave velocities among the measurements taken along the $\text{X}$ ($V_{p_x}$), $\text{Y}$ ($V_{p_y}$), and $\text{Z}$ ($V_{p_z}$) directions.
Accordingly, for the TZP, the average anisotropy ratio remains low, with a value of $1.52~\pm~0.4\%$. The low anisotropy ratio is consistent with the expected behavior of the sample, as zirconia ceramics manufactured under controlled industrial processes are typically isotropic and homogeneous in structure. Based on density, $\rho$, and average P-wave ($V_{p_{avg}}$),and S-wave ($V_{s_{avg}}$) velocities, the values of elasticity modulus can be evaluated as follows
\begin{equation}
    E = \rho V_{s_{avg.}}^2\frac{3V_{p_{avg.}}^2 - 4V_{s_{avg.}}^2}{V_{p_{avg.}}^2-V_{s_{avg.}}^2}
\end{equation}

\begin{equation}
    \nu = \frac{V_{p_{avg.}}^2 - 2V_{s_{avg.}}^2}{2(V_{p_{avg.}}^2 - V_{s_{avg.}}^2)}
\end{equation}

Under constant temperature conditions, the average Young’s modulus, $E$, was found to be $212.52~\pm~3.07~\text{GPa}$, while the average Poisson’s ratio, $\nu$, was approximately 0.30. These values are in good agreement with the manufacturer-provided reference values of  $E=210~\text{GPa}$ and $\nu=0.3$.

Similarly, to assess the robustness of the thermal conductivity measurements, a finite element analysis-based correction procedure was adopted. This approach was necessary to account for potential heat losses and non-ideal boundary conditions inherent in the experimental configuration. It should be noted that the initial attempt to calibrate the system using a quartz glass cube, similar to the material employed in the COC setup, was unsuccessful. During sample placement, the cubic quartz glass specimen failed, likely due to its brittle nature and the presence of micro-defects that made it susceptible to cracking under deviatoric stresses during the sample placement procedure. To overcome this limitation, a TZP cube was selected as the sample for the calibration process. 

\begin{figure}[htbp]
    \centering
    \includegraphics[width=0.65\linewidth]{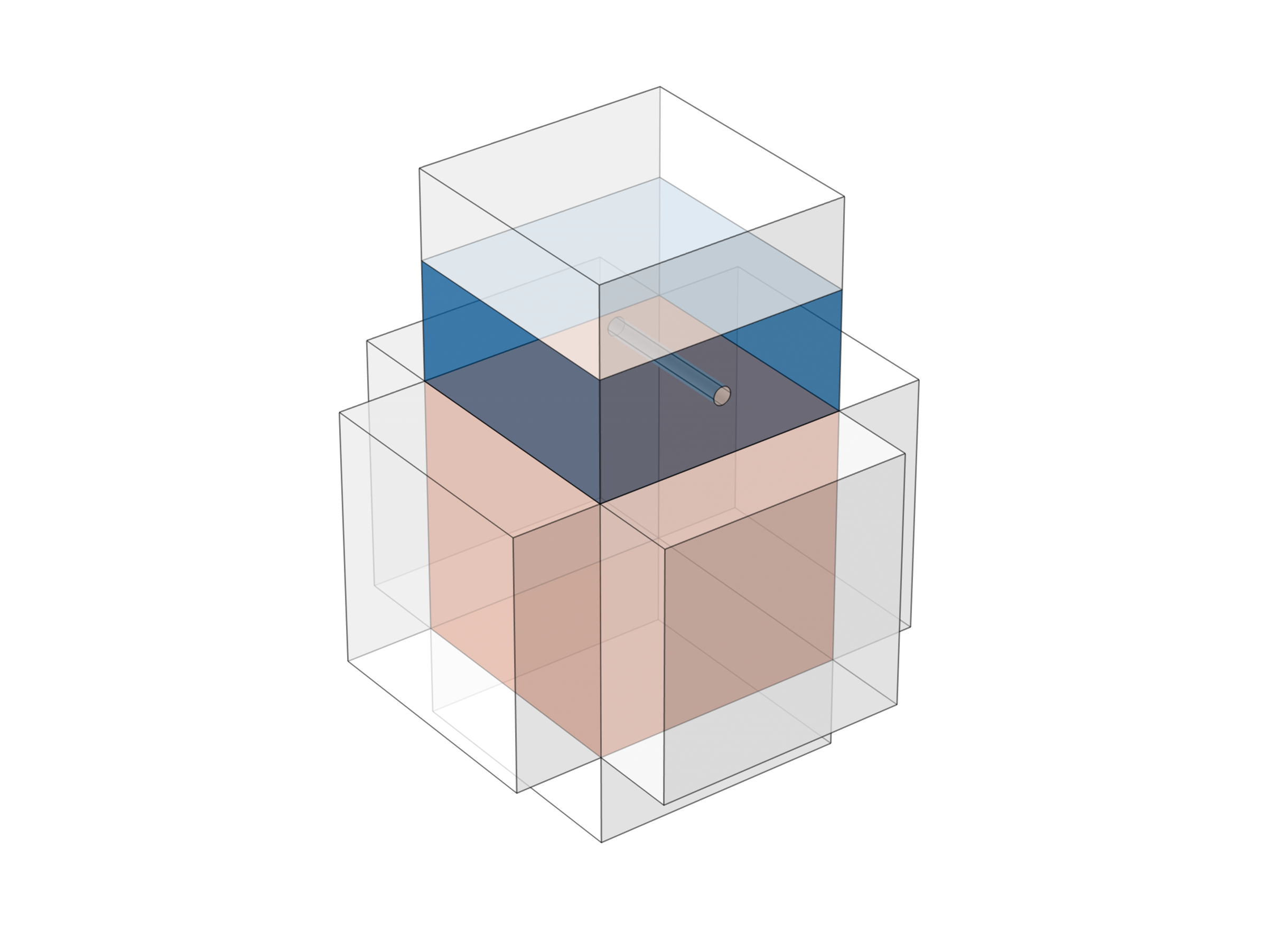}
    \caption{Numerical replica of the cubic press configuration used for thermal conductivity correction. The specimen (red) is positioned at the center, with the zirconia reference plate shown in blue. The end caps and pistons are displayed with a transparent rendering to highlight internal components.}
    \label{fig:WP_Comsol_Setup}
\end{figure}

The numerical model developed for this purpose is shown in Fig.~\ref{fig:WP_Comsol_Setup}. The sample is positioned between six loading/heating caps, while the reference material is placed at the top of the sample. For clarity, the main components are color-coded: the rock specimen is shown in red, the zirconia reference material in blue, and the end caps are rendered transparent to reveal the internal configuration. The FEM model incorporated the actual specimen dimensions measured during various stages of loading to ensure geometric accuracy. As in the previous case, the thermal conductivity of the specimen was iteratively adjusted in the simulation until the computed temperature distribution closely matched the experimental measurements.

\subsection{General remarks}

To account for unmodeled effects and to compensate for systematic bias in the setups, a temperature-dependent correction factor was introduced in each model. This approach is commonly applied during calibration and reference testing, whereby model parameters are tuned so that simulation results align with experimental observations. For example, in \cite{xing2014correction} the thermal conductivity of an insulation material was calibrated using reference experiments to minimise discrepancies between measured and simulated values.

In the present work, physics-based, temperature-dependent boundary conditions were imposed in the numerical models during calibration so that systematic bias would be removed. Because the same hardware, cabling, assembly, and boundary conditions are used in both the calibration and the measurement campaigns, these systematic effects are taken as already compensated. It is explicitly assumed that, after this correction, uncertainties in the boundary conditions and model inputs can be neglected; under this assumption, the remaining measurement error reflects only measurement scatter (temperature readings and displacement sensors). This assumption is valid provided that the calibration conditions are the same as those used in the actual measurements.

Across repeated calibration experiments, the maximum standard deviation of the measured temperatures under similar boundary conditions did not exceed $\pm 0.07~^{\circ}\text{C}$ for the COC setup and $\pm 0.12~^{\circ}\text{C}$ for the cubic press setup, indicating high repeatability. Accordingly, a maximum differential error for the estimated thermal conductivity is reported by propagating this observed measurement scatter through the conductivity calculation via standard chain-rule (first-order) derivatives, while other quantities are held at their nominal values. With this convention, and for a representative conductivity of 2.5~$\text{W}~\text{m}^{-1}~\text{K}^{-1}$ (typical of the tested rocks), the maximum differential error is approximately 5~\% for COC and 9~\% for the cubic press. The comparatively higher figure for the cubic press is attributed to the larger temperature scatter observed under identical boundary conditions.

 \section{Material}

\begin{figure}[htbp]
    \centering
    \begin{subfigure}[b]{0.45\textwidth}
        \centering
        \includegraphics[width=\textwidth, trim=0cm 10cm 0cm 5cm, clip=true]{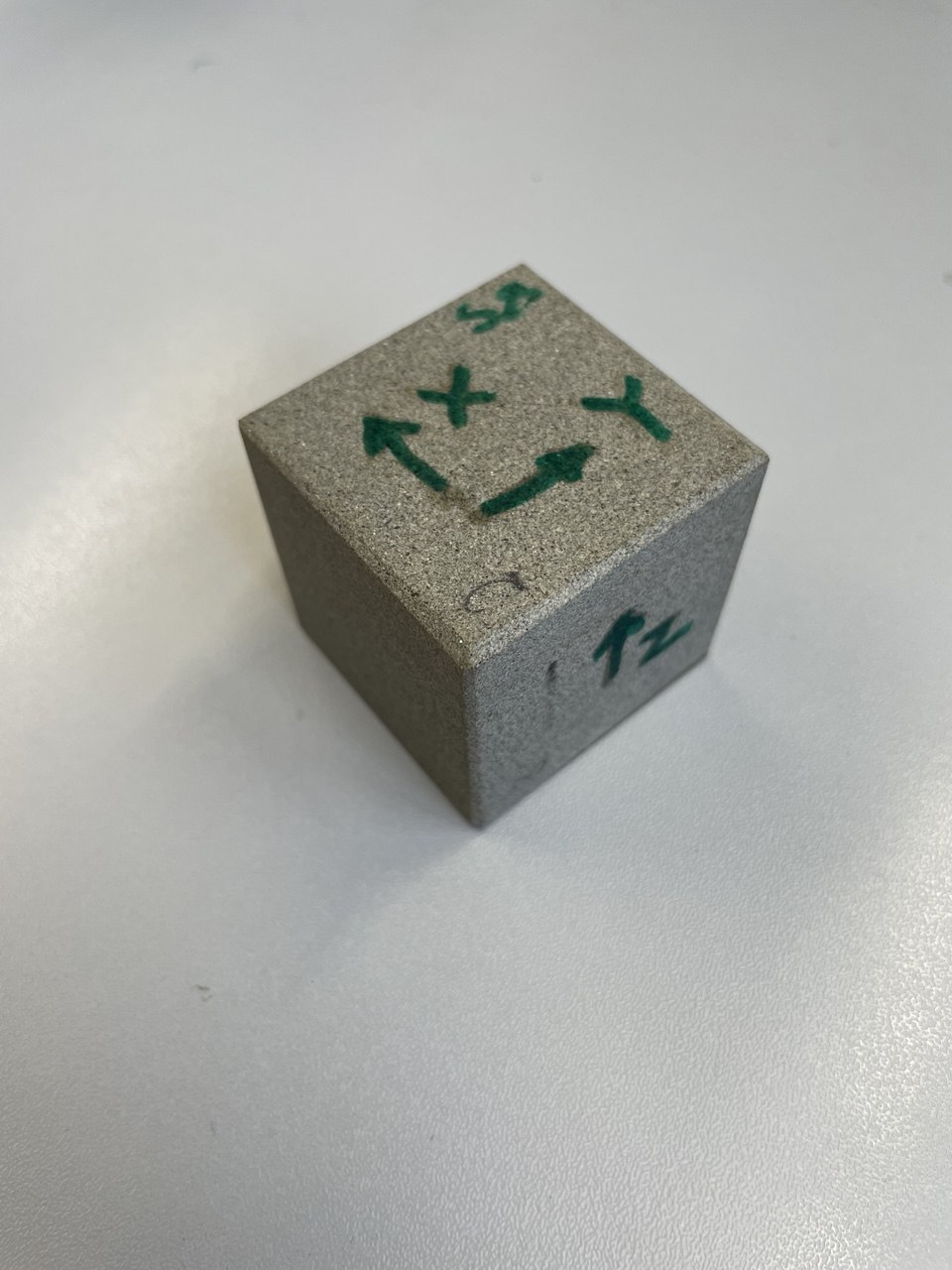}
        \subcaption{}
    \end{subfigure}
    \hfill
    \begin{subfigure}[b]{0.45\textwidth}
        \centering
        \includegraphics[width=\textwidth, trim=0cm 10cm 0cm 5cm, clip=true]{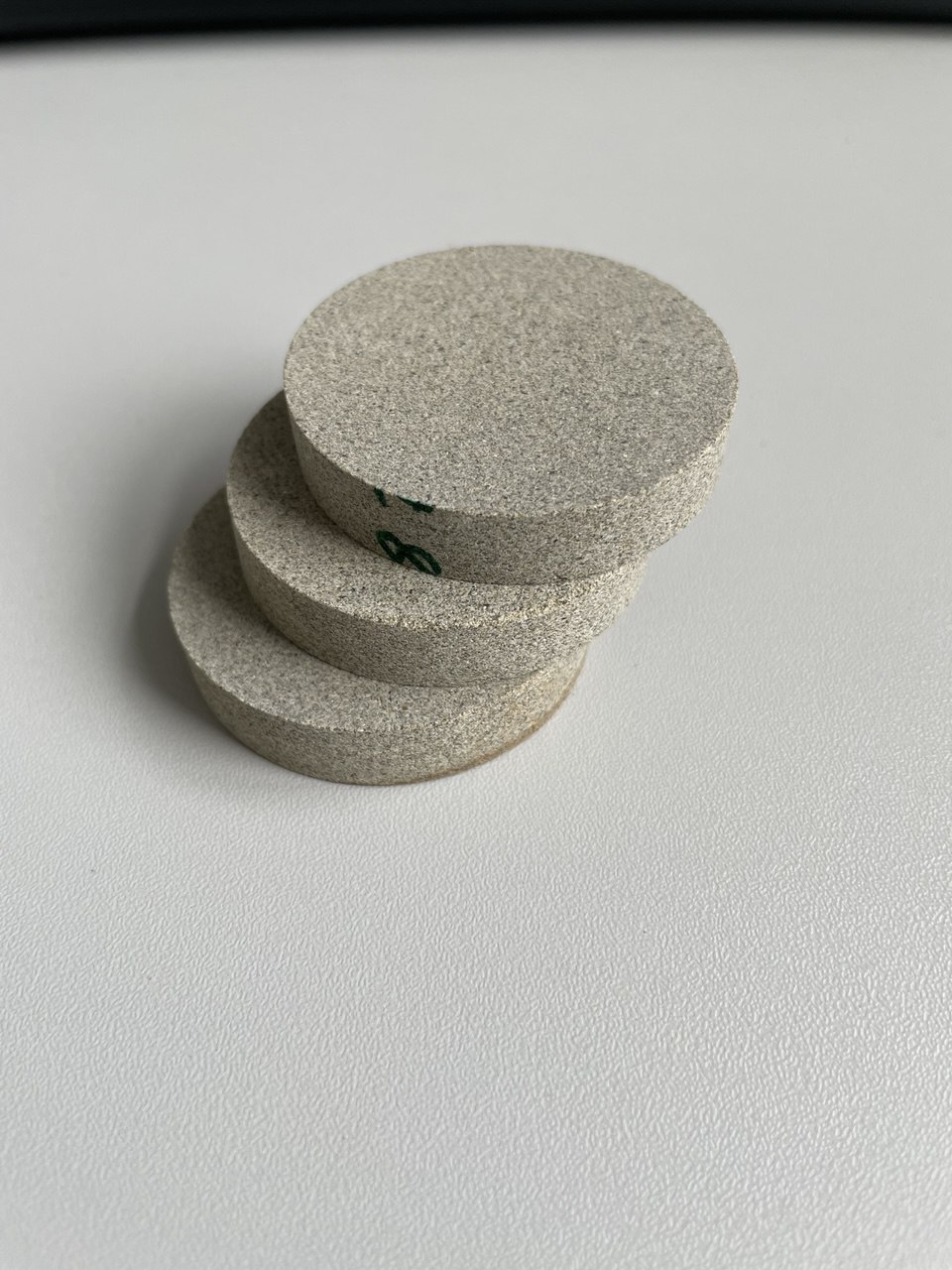}
        \subcaption{}
    \end{subfigure}
    \caption{Prepared (a) cubic, and (b) cylindrical samples used for ETC and wave velocity measurements. The marked coordinate axis on the cubic sample denotes the placement direction in the cube press.}
    \label{fig:s8_prepared_samples}
\end{figure}

 To investigate the combined effects of pressure and temperature on the effective thermal conductivity of rocks, two sandstone types, designated S\#8 and S\#12, were selected for this study. The rock specimens were extracted from sandstone outcrops sampled from the Frosinone Formation in the central Apennines, Italy~\citep{smeraglia2014tectonic}. The primary distinction between these two sandstone types lies in their porosity values and the distance from the nearby fault where they were collected~\citep{smeraglia2014tectonic}. Sandstone S\#8 has a porosity of 8.4~\%, while S\#12 has a porosity of 5.68~\%. From each type, a cubic sample with a side length of 43mm was prepared for the simultaneous measurement of ETC and three-dimensional wave velocity in the cube press. Additionally, a series of cylindrical samples was prepared for uniaxial ETC measurements using the COC device. The example prepared samples for the S\#8 are shown in Fig.~\ref{fig:s8_prepared_samples}. To prevent damage at the corner edges and reduce stress concentrations, the edges of the samples were slightly flattened. For the cylindrical samples, a 5~cm core was first drilled from the outcrop and then sliced into disks with a height of 1~cm.

 In the initial phase, to further characterize the mineralogy and microstructure of the sandstone samples, an array of X-ray fluorescence (XRF), X-ray diffraction (XRD), thin section petrography, and Scanning Electron Microscopy (SEM) has been conducted. Using the XRF method, the major elemental composition of the sandstone was determined on air-dried, crushed, and milled samples in the form of pressed tablets. The results are given in Table.~\ref{tab:TabSandStoneXRF}.

\begin{table}[htbp]
\centering
\caption{XRF analysis (mass~\%) of the samples}
\begin{tabular}[t]{l c c c c c c c}
\toprule
Sample name & $SiO_2$ & $Al_2O_3$ & $MgO$ & $Fe_2O_3$ & $CaO$ & $K_2O$ & $Na_2O$\\
\midrule
S\#8 & 53.107 & 13.098 & 4.637 & 4.927 & 18.223	& 2.704 & 1.783\\
S\#12 & 48.023 & 11.645 & 4.267 & 5.118 & 25.342 & 2.281 & 1.654\\
\bottomrule\\
\end{tabular}
\label{tab:TabSandStoneXRF}
\end{table}

\begin{figure}
    \centering
    \includegraphics[width=0.65\linewidth]{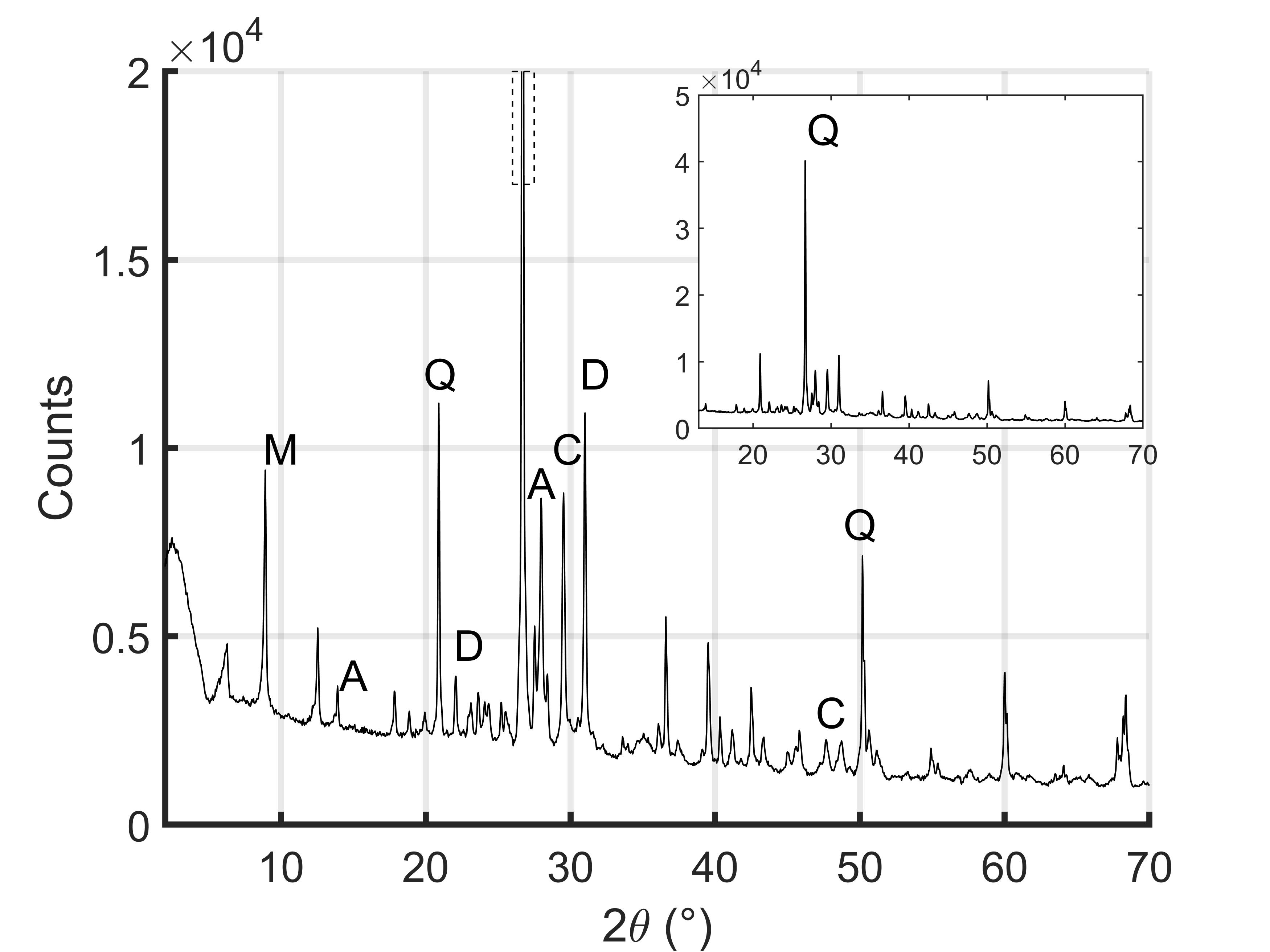}
\caption{XRD phase analysis of sandstone. Peak labels correspond to minerals as follows: Q – Quartz, D – Dolomite, C – Calcite, A – Albite, and M – Mica.}
    \label{fig:S8_XRD}
\end{figure}

Accordingly, as shown in Fig.~\ref{fig:S8_XRD}, an example of XRD patterns from powdered sandstone samples is illustrated. The XRD results, further backed by the XRF analysis, revealed that quartz is the dominant constituent mineral in the studied sample, with other minerals including dolomite, albite, and calcite. Rietveld analysis for S\#8 shows that quartz, calcite, dolomite, and albite constitute approximately 40~\%, 11~\%, 17~\%, and 32~\% of the rock volume, respectively. Similarly, S\#12 consists of about 37~\% quartz, 20~\% calcite, 17~\% dolomite, and 26~\% albite.

\begin{figure}[htbp]
    \centering
    \begin{subfigure}{0.45\textwidth}
        \includegraphics[width=\textwidth]{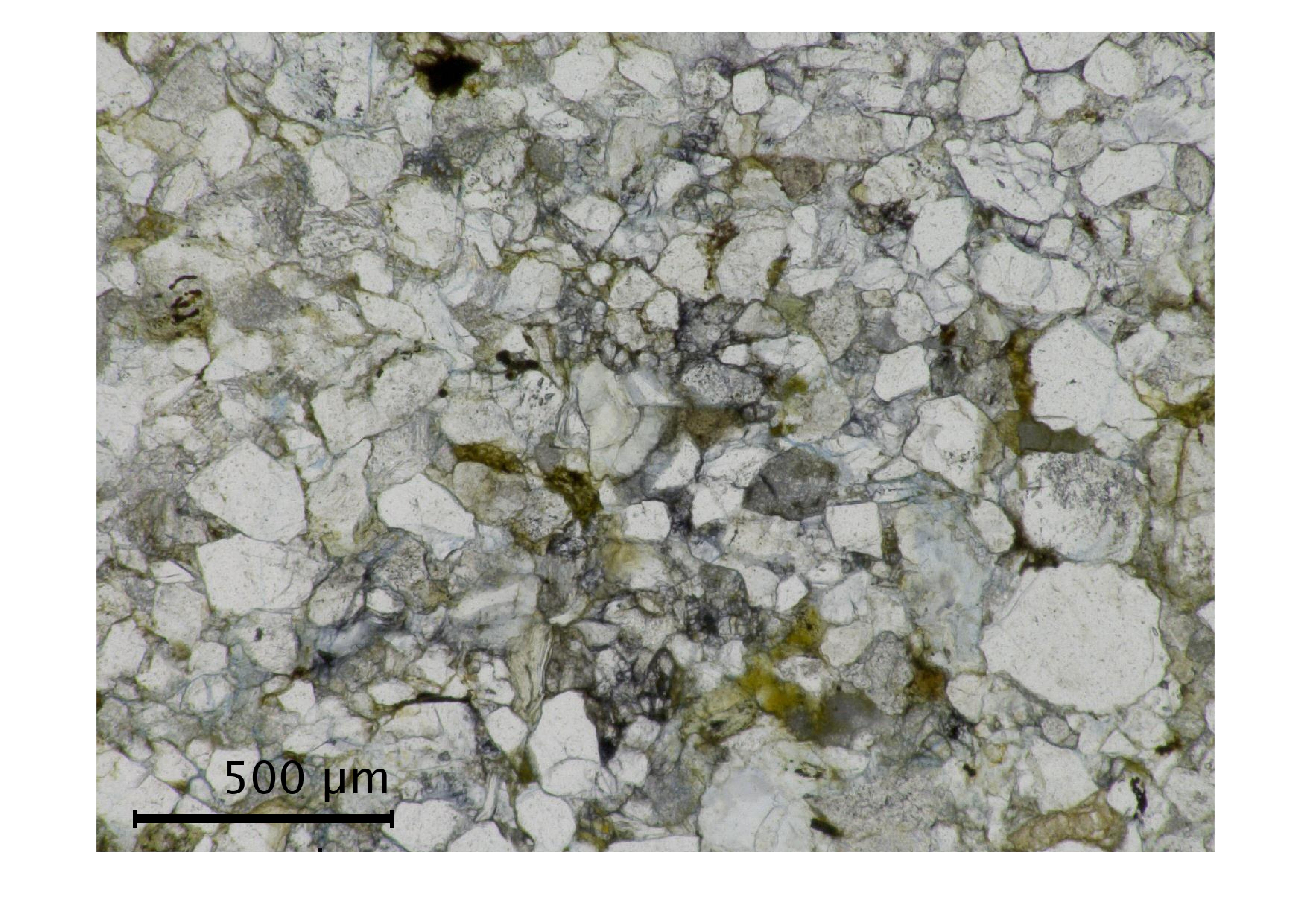}
        \subcaption{}
    \end{subfigure}
    \begin{subfigure}{0.45\textwidth}
        \includegraphics[width=\textwidth]{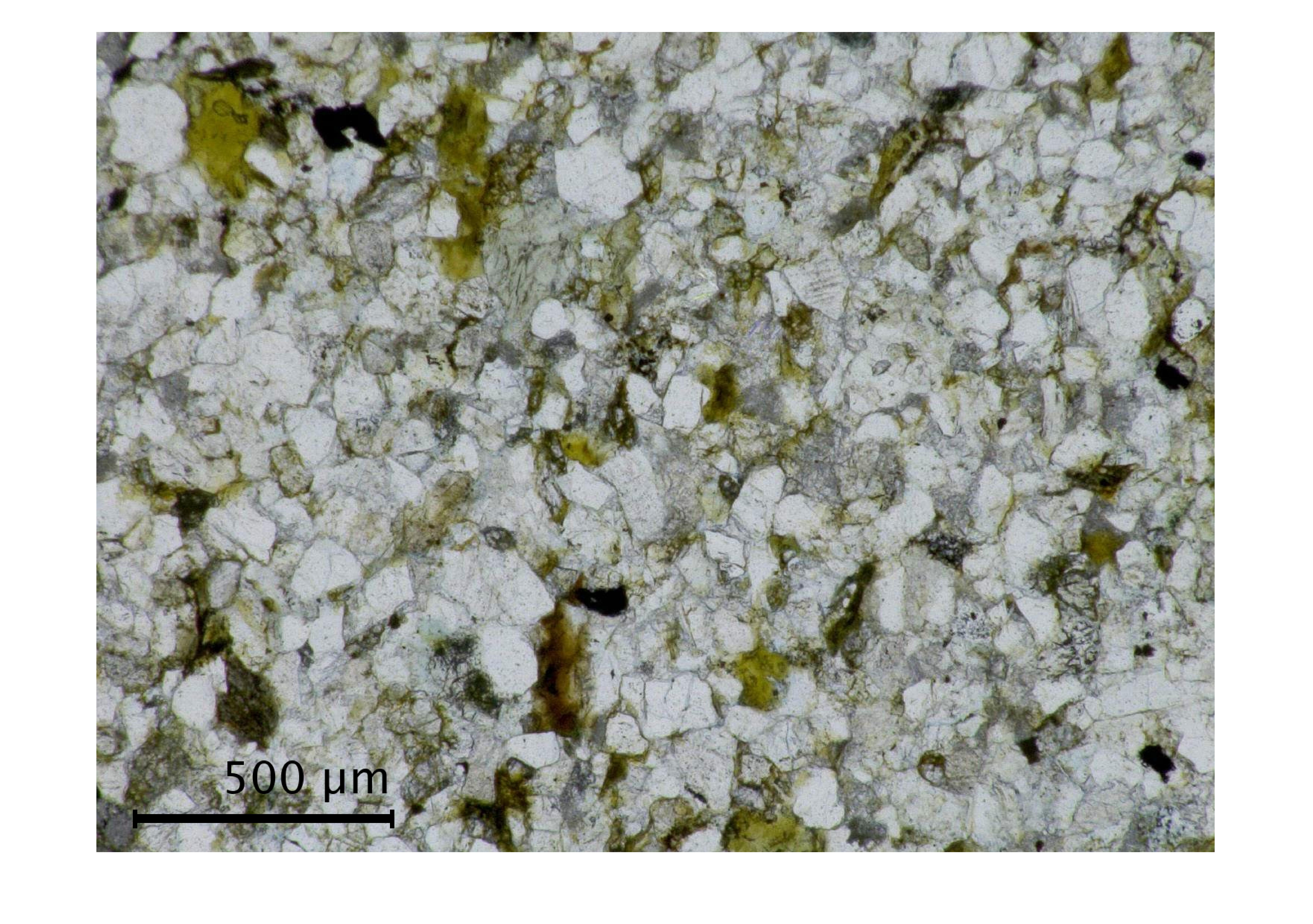}
        \subcaption{}
    \end{subfigure}
    \caption{Thin sections in the x–y principal plane viewed under plane-polarized light: (a) S\#8 (left) and (b) S\#12 (right). Scale bar = 0.5 mm.}
    \label{fig:S8_S12_Overall}
\end{figure}

\begin{figure}[htbp]
  \centering
  \begin{subfigure}[b]{0.3\textwidth}
    \includegraphics[width=\textwidth]{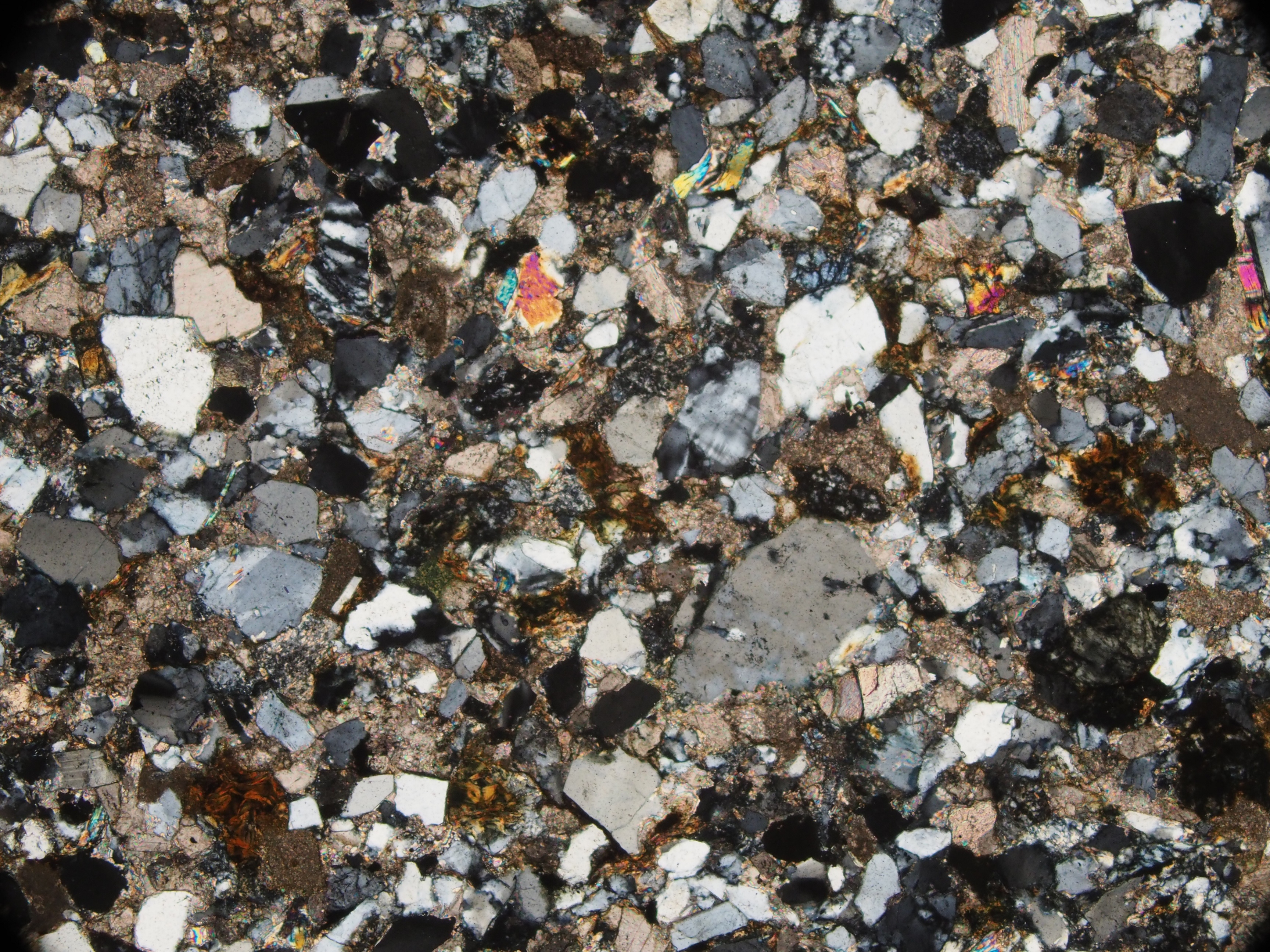}
    \subcaption{}
    \label{fig:BasicETCModels}
  \end{subfigure}
  \begin{subfigure}[b]{0.3\textwidth}
    \includegraphics[width=\textwidth]{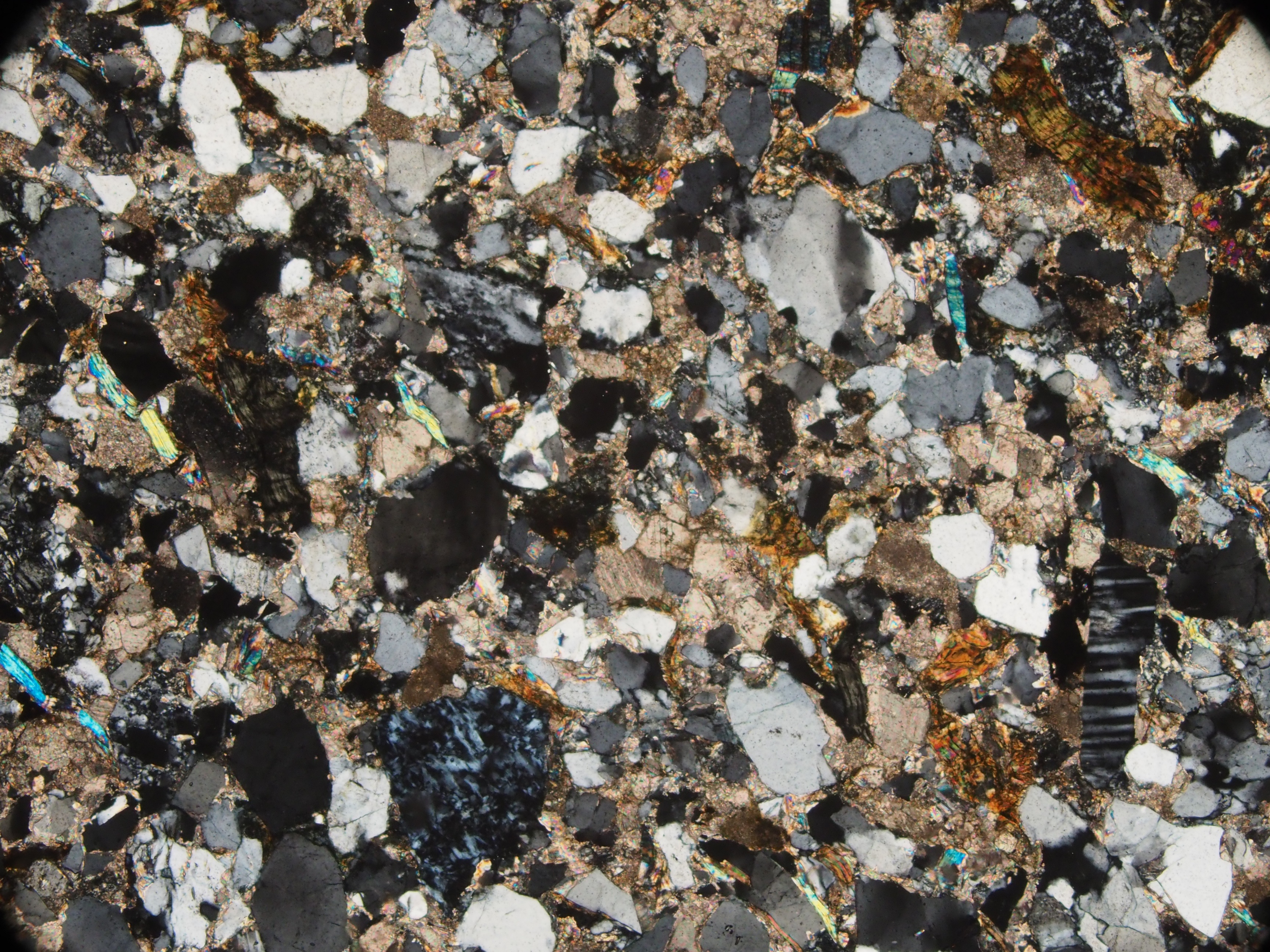}
    \subcaption{}
    \label{fig:CmbinedETCModel}
  \end{subfigure}
    \begin{subfigure}[b]{0.3\textwidth}
    \includegraphics[width=\textwidth]{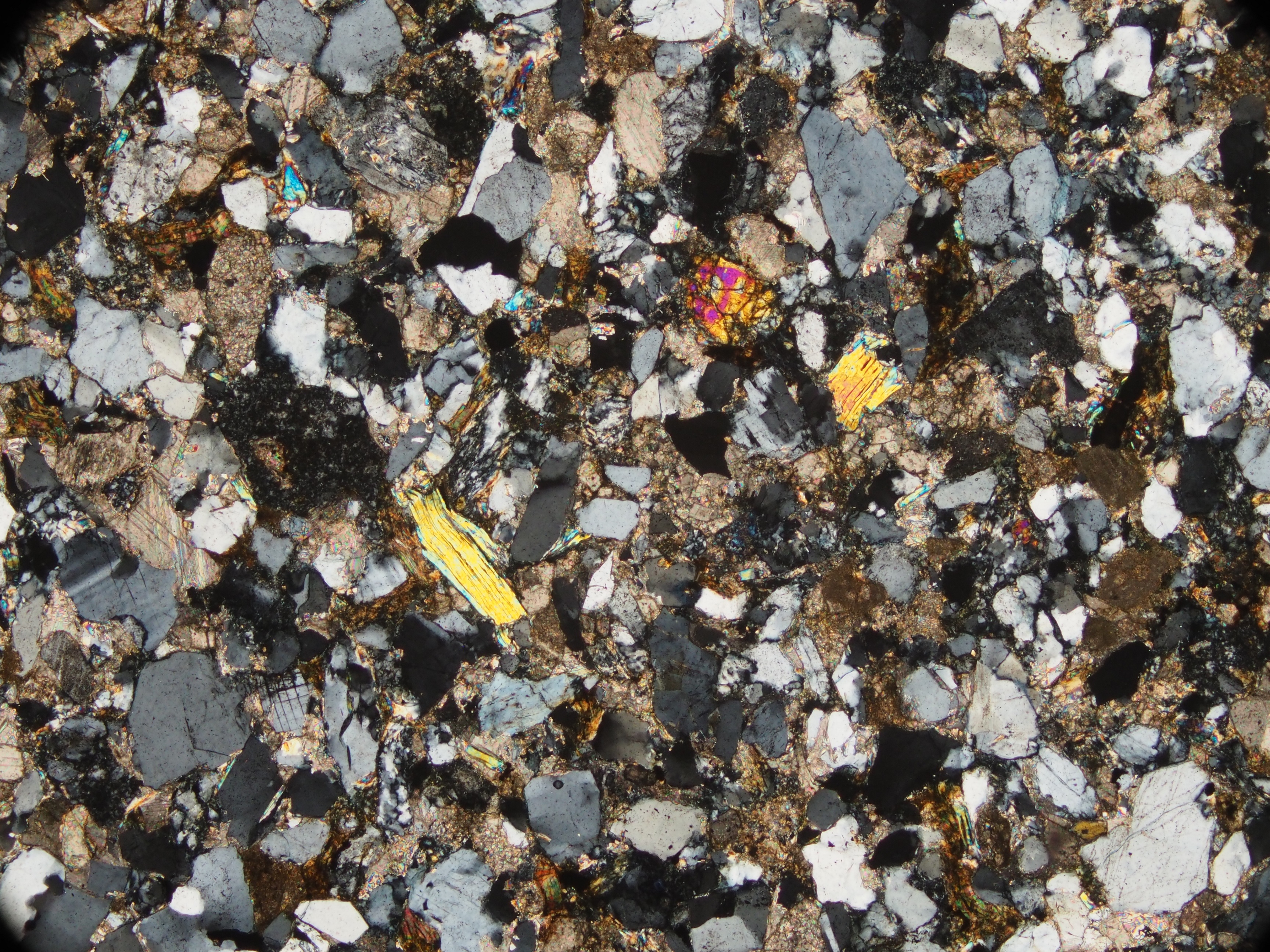}
    \subcaption{}
  \end{subfigure}
  \caption{Detailed thin section image of S\#8 under cross-polarized light in three main principal planes of (a) $x$--$y$, (b) $x$--$z$, and (c) $y$--$z$}
      \label{fig:xpl_xyz_s8}
\end{figure}

For each sandstone, three thin sections were prepared parallel to the $x$--$y$, $x$--$z$, and $y$--$z$ planes. The coordinate system is aligned with that shown in Fig.~\ref{fig:s8_prepared_samples}(a). The thin sections were prepared using a blue dye technique to facilitate the distinction between porosity and grains. An overall scaled image is first presented for the $x$--$y$ section to provide a general view of the mineral distribution (Fig.~\ref{fig:S8_S12_Overall}). As can be seen, the grains are usually angular to sub-angular and vary in size. In this regard, various automated image-processing approaches were initially tested to identify grain boundaries; however, the combination of limited image resolution and very low porosity prevented reliable detection. As a result, individual grains were manually measured using ImageJ, with the approximate equivalent circular diameter ranging from 47~$\mu m$ to 220~$\mu m$ for S\#8, and from 40~$\mu m$ to 136~$\mu m$ for S\#12. For each orientation, including the $x$--$y$ plane, this is followed by a higher-magnification image highlighting finer structural details. Given the broadly similar mineralogical composition of the two sandstones, detailed microstructural observations under cross-polarized light are presented only for S\#8 (Fig.~\ref{fig:xpl_xyz_s8}) as a representative example. The primary visual distinction between S\#8 and S\#12 lies in the more pronounced layering observed in S\#12 parallel to $x$--$y$ plane.

\begin{figure}[htbp]
    \centering
    \begin{subfigure}[b]{0.45\textwidth}
        \centering
        \includegraphics[width=\textwidth]{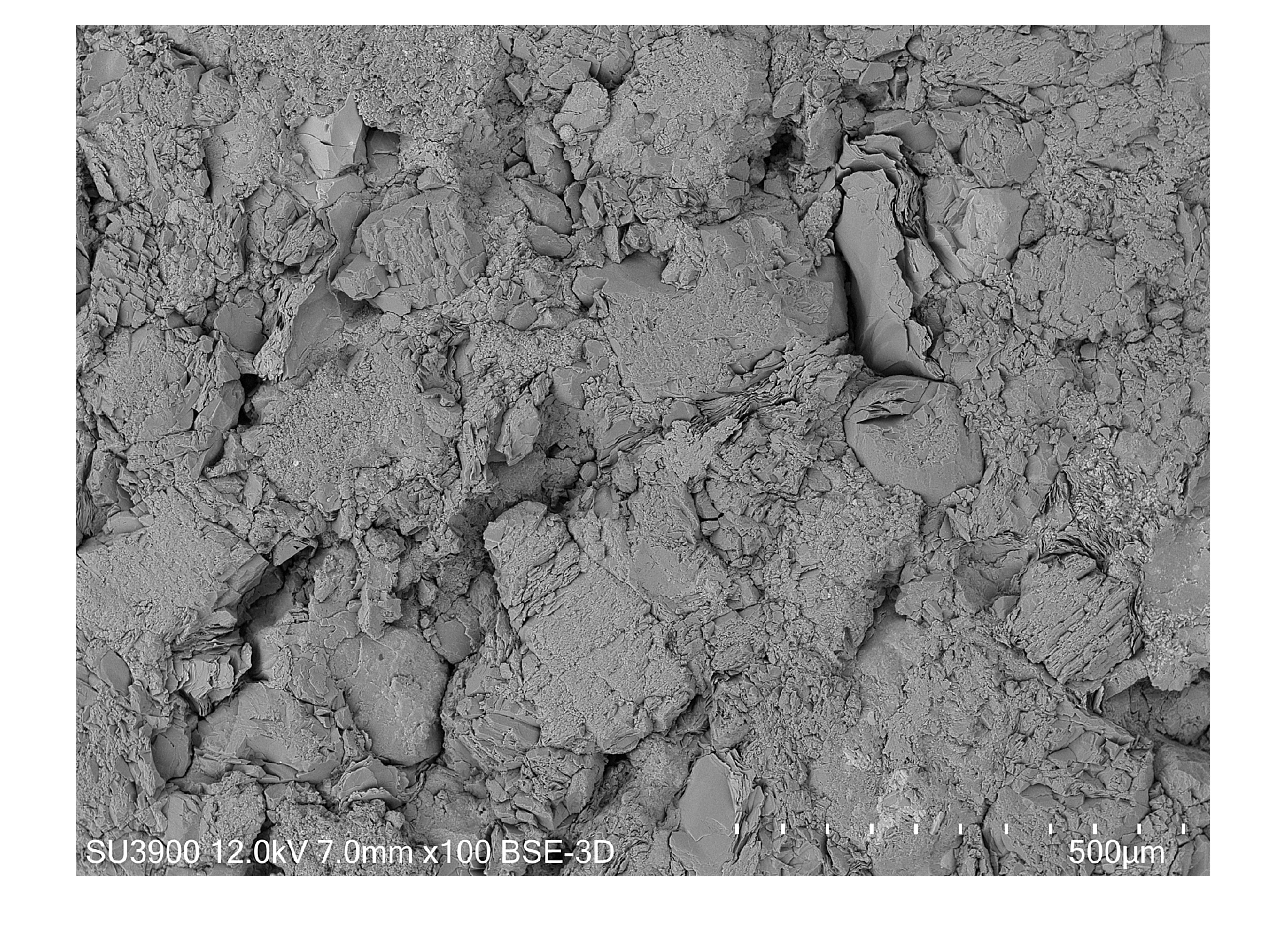}
        \subcaption{}
    \end{subfigure}
    \hfill
    \begin{subfigure}[b]{0.45\textwidth}
        \centering
        \includegraphics[width=\textwidth]{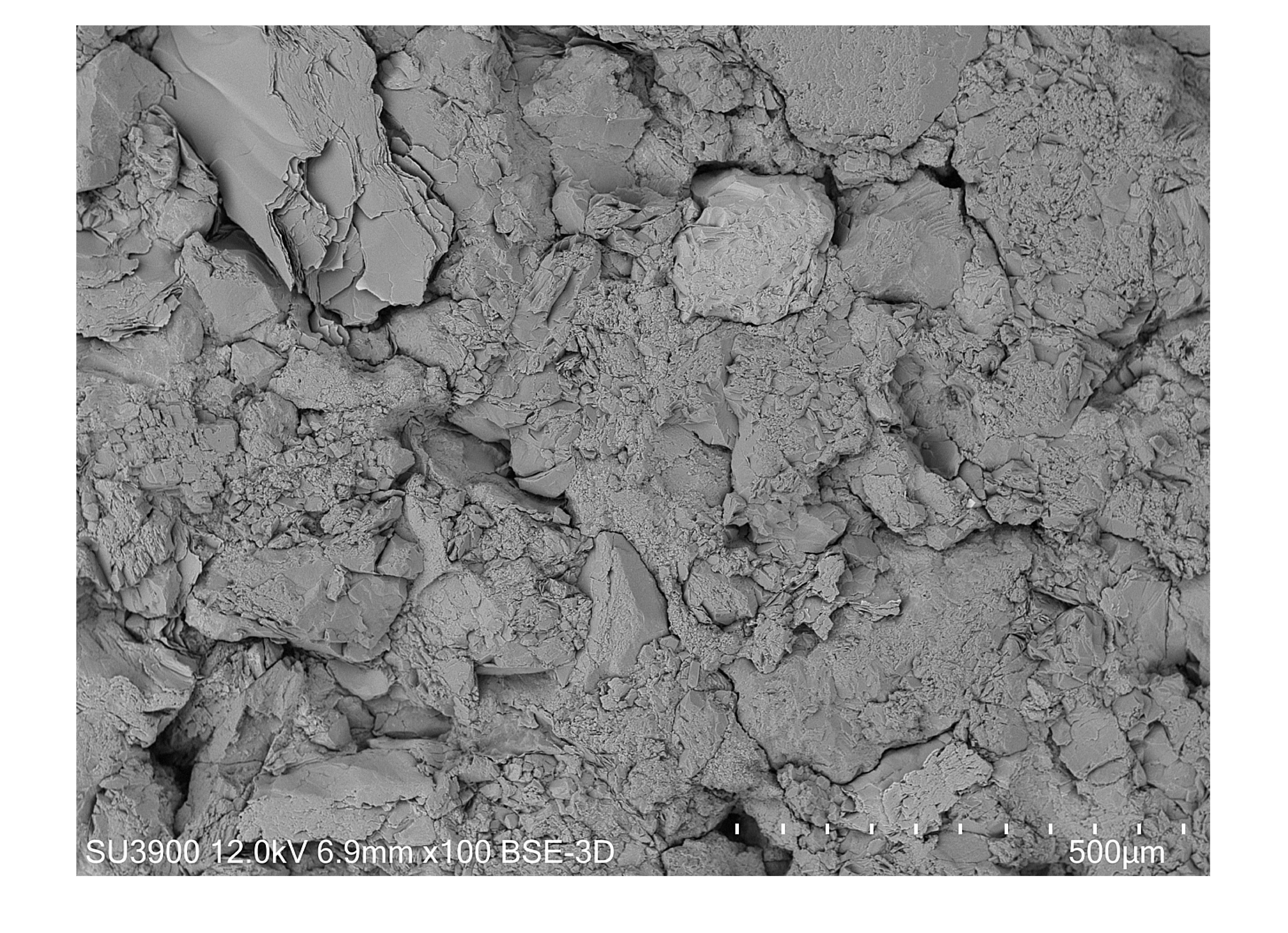}
        \subcaption{}
    \end{subfigure}
    \caption{SEM image of the sandstone microstructure.}
    \label{fig:SEM_S8}
\end{figure}

In the XPL images of the thin sections taken in the three principal directions for the S\#8, no preferred alignment of the grains can be observed, indicating a low degree of anisotropy in the sample. The observed mineral phases are quartz, feldspar, carbonates, and mica, mainly muscovite, which occur in a cement of fine-grained carbonate minerals. Carbonate cement fills most of the pore space and is easily identified because of its high birefringence, high relief, and prominent cleavage. Larger clasts of calcite, exhibit higher-order interference colors, and show characteristic rhombohedral cleavage and twinning planes. Quartz is the dominant clast and appears in shades of white to light gray due to its low birefringence. Larger quartz grains often show undulose extinction. Feldspar is present in both potassium-rich and sodium-rich varieties: potassium feldspar is recognized by its cross-hatched twinning typical of microcline, whereas sodium-rich plagioclase feldspar displays polysynthetic twinning. White mica is present in the thin sections as thin, platy, or flake-like grains, reflecting their perfect basal cleavage. Under cross-polarized light, they exhibit strong birefringence, producing bright and often colorful interference colors. To further illustrate this microstructural feature, representative SEM images of the studied sample is presented in Fig.~\ref{fig:SEM_S8}. Prepared samples for SEM analysis were placed in an ultrasonic bath to remove surface debris and subsequently gold-coated prior to imaging. As can be seen, a significant portion of the porosity is
observed in the form of narrow inter-granular and intra-granular cracks rather than large, connected pore spaces.

\section{Results}

This section presents the experimental findings for the two sandstone samples. Results are organized by sample and measurement type, with elastic wave velocity measurements presented first, followed by thermal conductivity data. The experiments were conducted under controlled pressure and temperature conditions using two complementary setups: the COC apparatus and the cubic press. The COC experiments provide baseline data at relatively low pressures, while the cubic press experiments extend the investigation to higher pressures and temperatures relevant to in-situ conditions.

In the COC setup, measurements were performed at the pressure level of 12~MPa, with the temperature increased stepwise from $50~^\circ\text{C}$ to $150~^\circ\text{C}$ in $25~^\circ\text{C}$ increments. In the cubic press, measurements were conducted at three pressure levels of 12, 50, and 100 MPa, with the temperature increased from $50~^\circ\text{C}$ to $250~^\circ\text{C}$ in $50~^\circ\text{C}$ increments. Results from the COC setup serve as reference points for the interpretation of the more advanced cubic press measurements.

\begin{figure}[htbp]
    \centering
    \includegraphics[width=\textwidth]{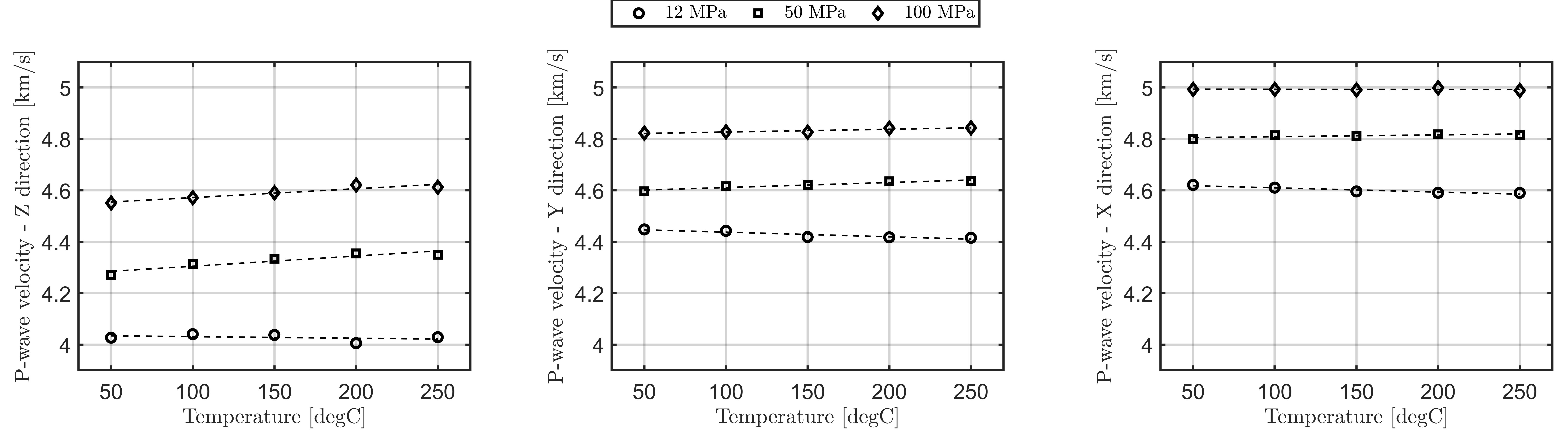}
    \vspace{0.2em}

    \begin{tabular}{@{}ccc@{}}
    \makebox[0.33\textwidth][c]{(a)} &
    \makebox[0.32\textwidth][c]{(b)} &
    \makebox[0.31\textwidth][c]{(c)}
    \end{tabular}
    
    \vspace{0.5em}
    \includegraphics[width=\textwidth]{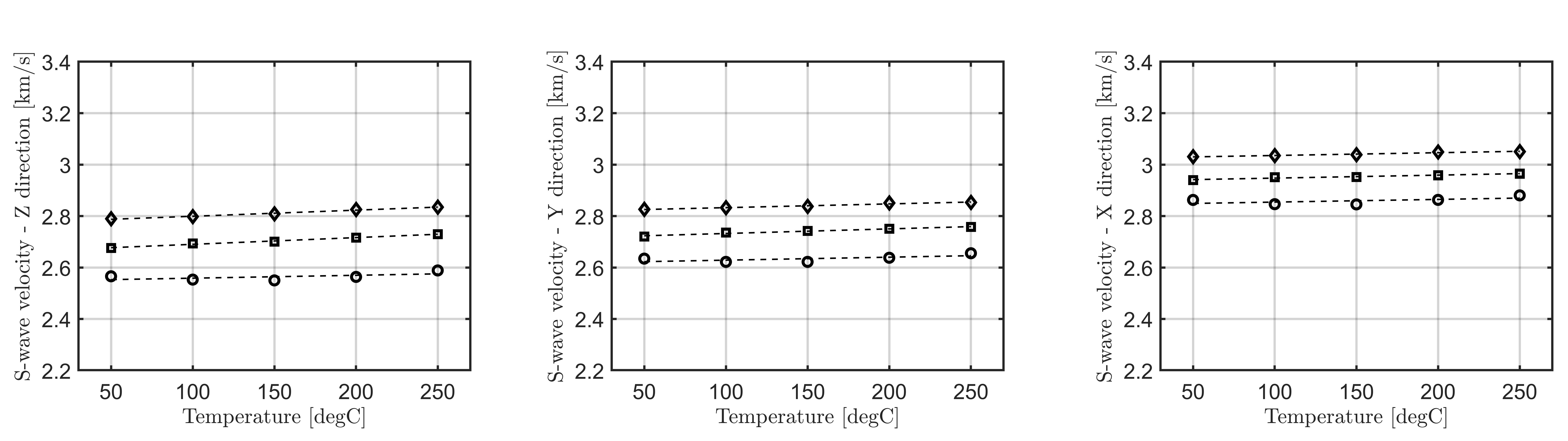}
    \vspace{0.2em}

    \begin{tabular}{@{}ccc@{}}
    \makebox[0.33\textwidth][c]{(d)} &
    \makebox[0.32\textwidth][c]{(e)} &
    \makebox[0.31\textwidth][c]{(f)}
    \end{tabular}
    \caption{P-wave (top panels) and average polarized S-wave velocities (bottom panels) measured along the three principal orthogonal directions for sandstone samples S\#8 as a function of applied temperature and confining stress.}
    \label{fig:S8_overall}
\end{figure}

For S\#8, P-wave velocities measured along the three principal orthogonal directions are presented in Fig.\ref{fig:S8_overall} (a), (b), and (c), respectively, showing their variation with applied temperature and stress. Similarly, the corresponding averages of polarized S-wave velocities in the x–y, x–z, and y–z planes are shown in Fig.\ref{fig:S8_overall}, (d), (e), and (f), respectively, for the same pressure–temperature conditions.

\begin{figure}[htbp]
    \centering
    \includegraphics[width=\textwidth]{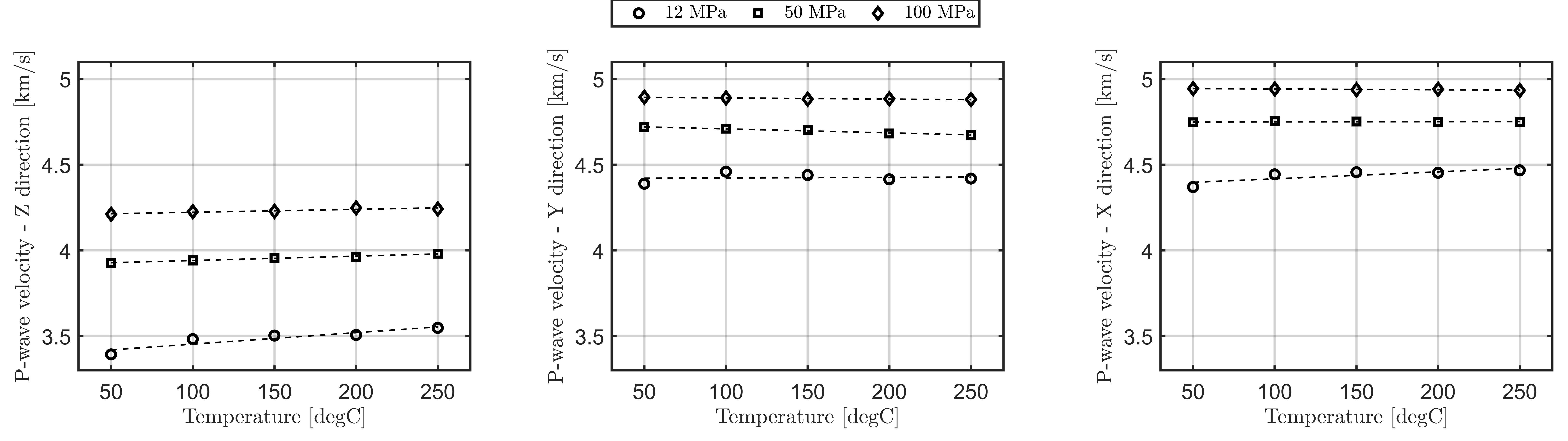}
    \vspace{0.2em}

    \begin{tabular}{@{}ccc@{}}
    \makebox[0.33\textwidth][c]{(a)} &
    \makebox[0.32\textwidth][c]{(b)} &
    \makebox[0.31\textwidth][c]{(c)}
    \end{tabular}
    
    \vspace{0.5em}
    \includegraphics[width=\textwidth]{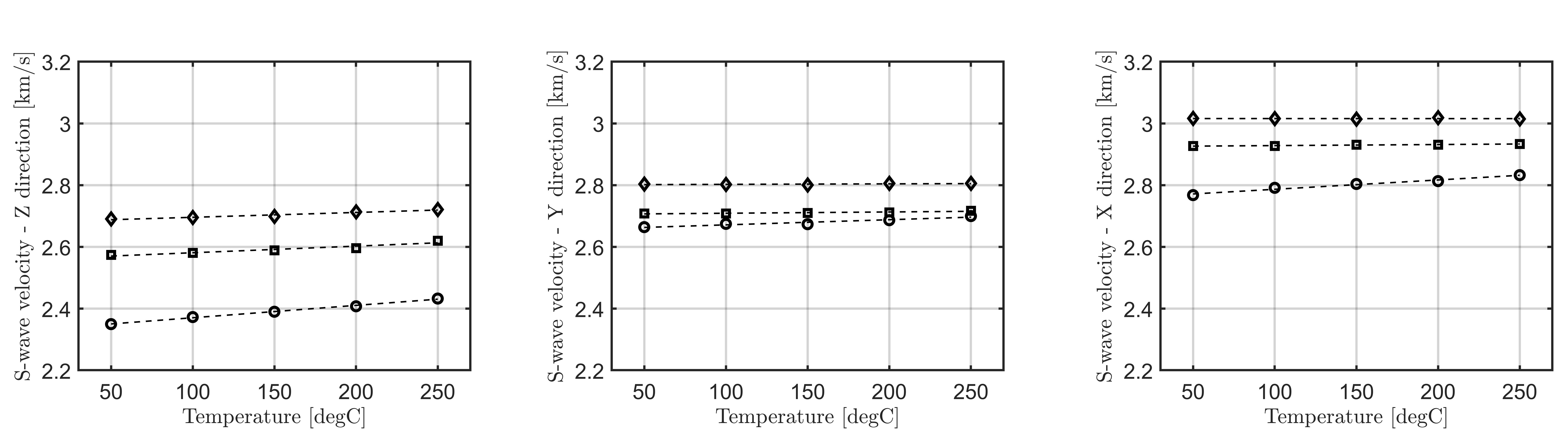}
    \vspace{0.2em}

    \begin{tabular}{@{}ccc@{}}
    \makebox[0.33\textwidth][c]{(d)} &
    \makebox[0.32\textwidth][c]{(e)} &
    \makebox[0.31\textwidth][c]{(f)}
    \end{tabular}
    \caption{P-wave (top panels) and average polarized S-wave velocities (bottom panels) measured along the three principal orthogonal directions for sandstone samples S\#12 as a function of applied temperature and confining stress.}
    \label{fig:S12_overall}
\end{figure}

For S\#12, P- and S-wave velocity measurements were conducted under identical experimental conditions. The results, presented in Figs.~\ref{fig:S12_overall}, illustrate the directional dependence of elastic wave velocities and their evolution with increasing temperature and stress.

\begin{figure}[htbp]
    \centering
    \begin{subfigure}[t]{0.4\textwidth}
        \centering
        \includegraphics[width=\textwidth]{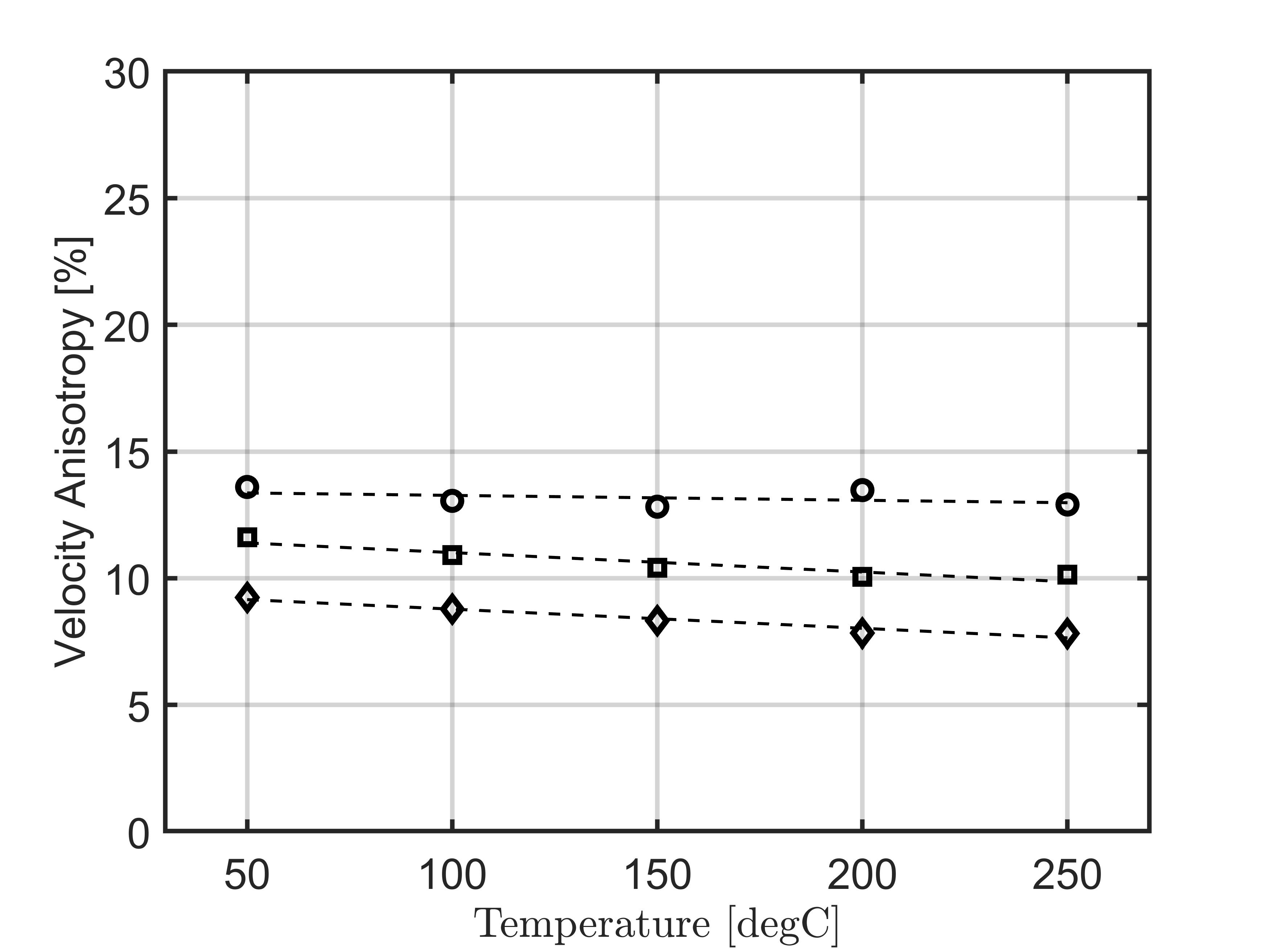}
        \subcaption{}
    \end{subfigure}
    \begin{subfigure}[t]{0.4\textwidth}
        \centering
        \includegraphics[width=\textwidth]{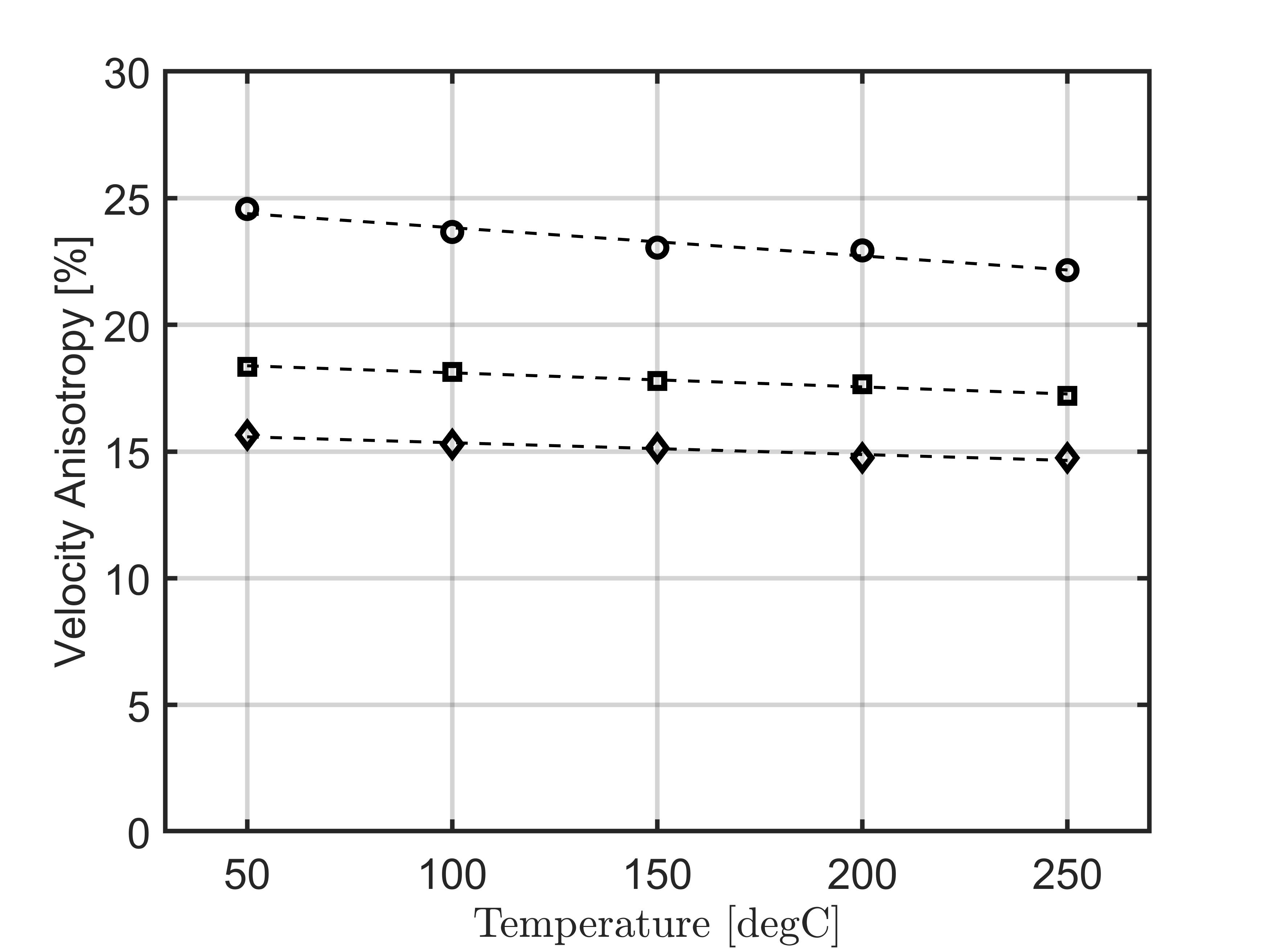}
        \subcaption{}
    \end{subfigure}
    \begin{subfigure}[t]{0.15\textwidth}
        \centering
        \includegraphics[width=\textwidth]{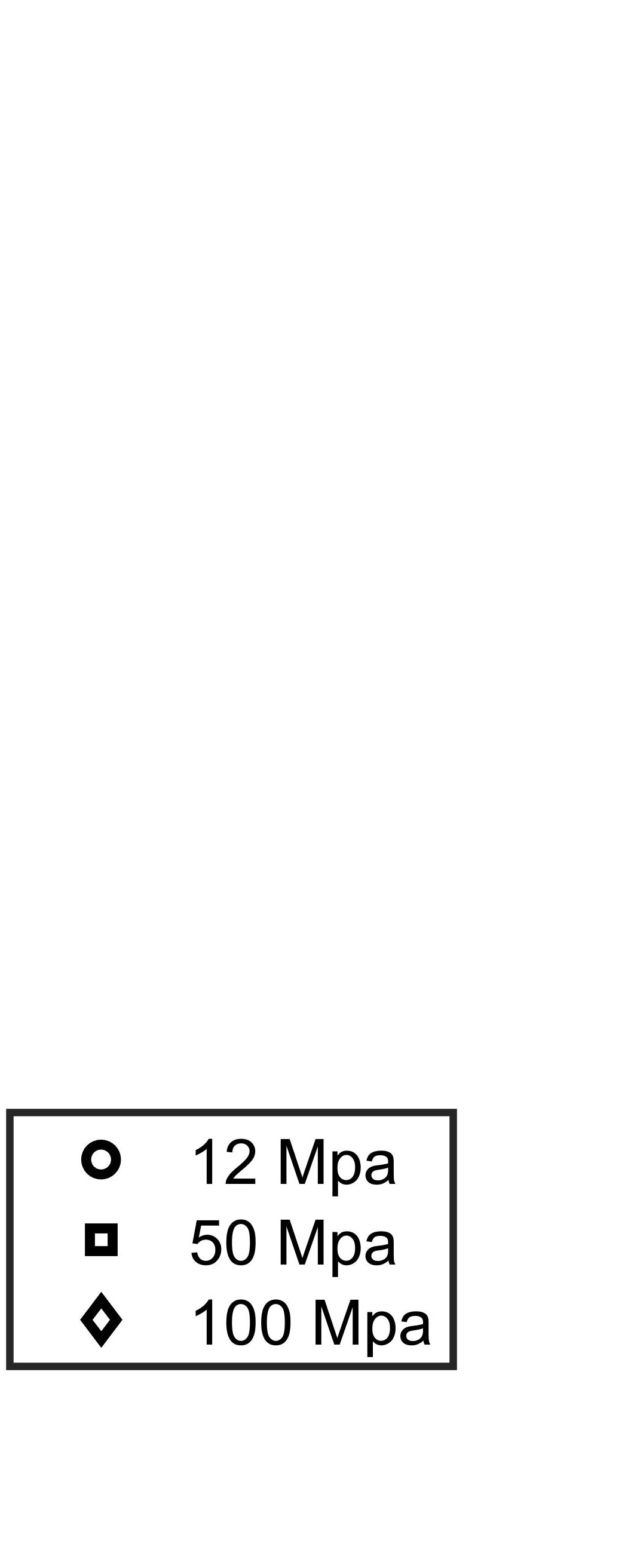}
    \end{subfigure}
    \caption{Evolution of elastic anisotropy in sandstone samples S\#8 (left) and S\#12 (right) under varying pressure and temperature conditions.}
    \label{fig:anisotropy_s12_s8}
\end{figure}

 Based on initial observations, both sandstone samples exhibit similar trends in the evolution of wave velocities with varying pressure and temperature. The first notable difference is the comparatively higher anisotropy for S\#12, a behavior that is less pronounced in S\#8. This can be attributed to the more prominent layered structure observed in S\#12. In this regard, the P-wave anisotropy for both samples, as defined in Eq.~(\ref{eq:anisotropy}), is illustrated in Fig.~\ref{fig:anisotropy_s12_s8}. In the first step of the experiment, the P-wave anisotropy for S\#8 is approximately 15~\%, whereas S\#12 exhibits a higher anisotropy of around 25~\%

 Regarding the effects of pressure and temperature, wave velocities increase with applied pressure, reflecting the expected stiffening of the rock framework. In contrast, the influence of temperature is less straightforward: only a very slight decrease or, in some cases, an increase in wave velocities is observed. This behavior deviates from the generally expected decrease in rock wave velocities with increasing temperature. The observed decrease in anisotropy and the slight increase in wave velocities with rising temperature, although uncommon, have also been reported in previous studies~\citep{scheu2006temperature}. This behavior can be attributed to the low porosity and specific pore structure of the studied sandstones. At elevated pressures, the applied confinement likely restricts thermal expansion of the mineral grains, increasing the rock’s elastic stiffness and, consequently, enhancing the wave velocities.

\begin{figure}[htbp]
    \centering
    \begin{subfigure}[t]{0.4\textwidth}
        \centering
        \includegraphics[width=\textwidth]{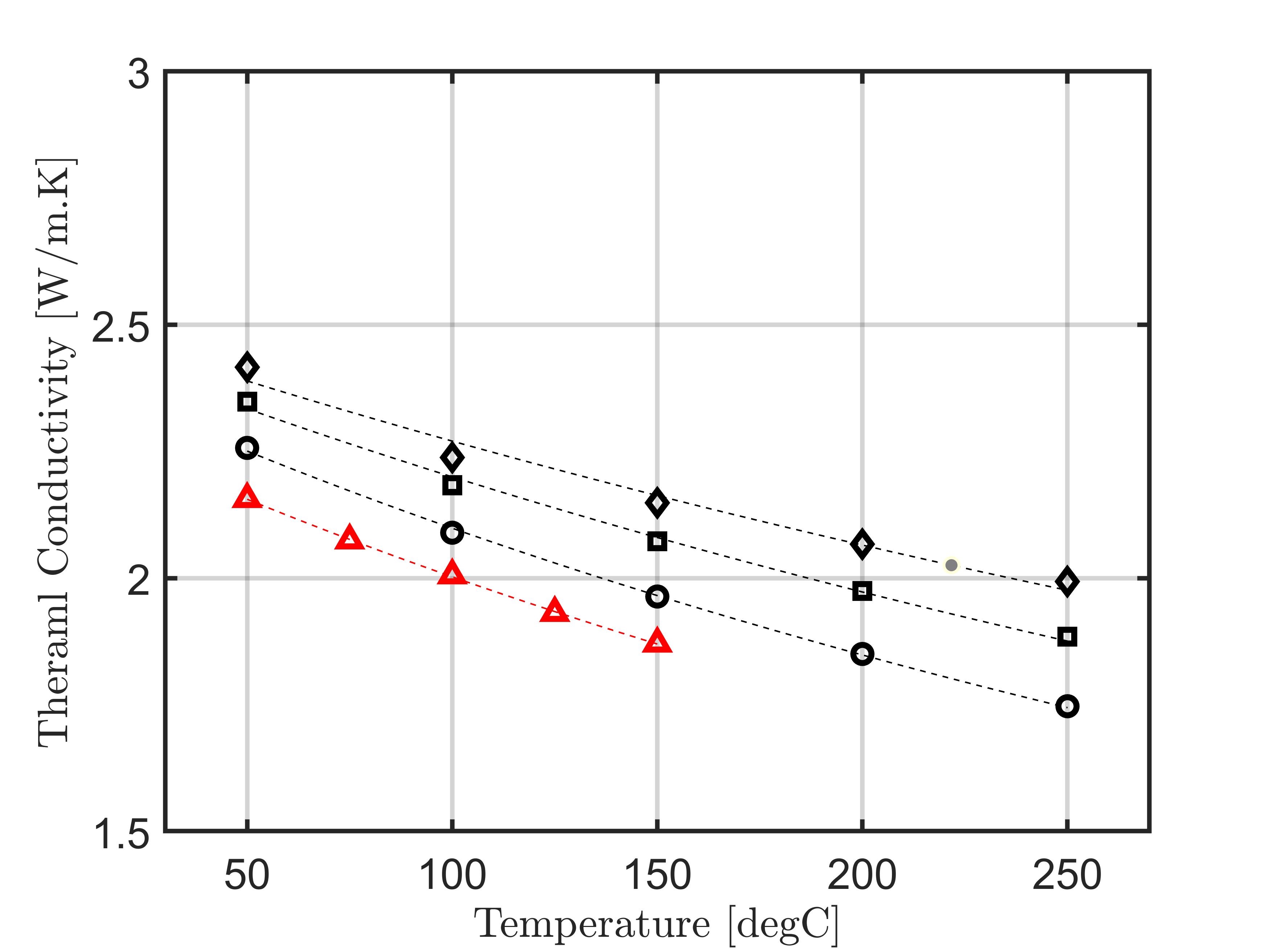}
        \subcaption{\label{fig:TC_S8}}
    \end{subfigure}
    \begin{subfigure}[t]{0.4\textwidth}
        \centering
        \includegraphics[width=\textwidth]{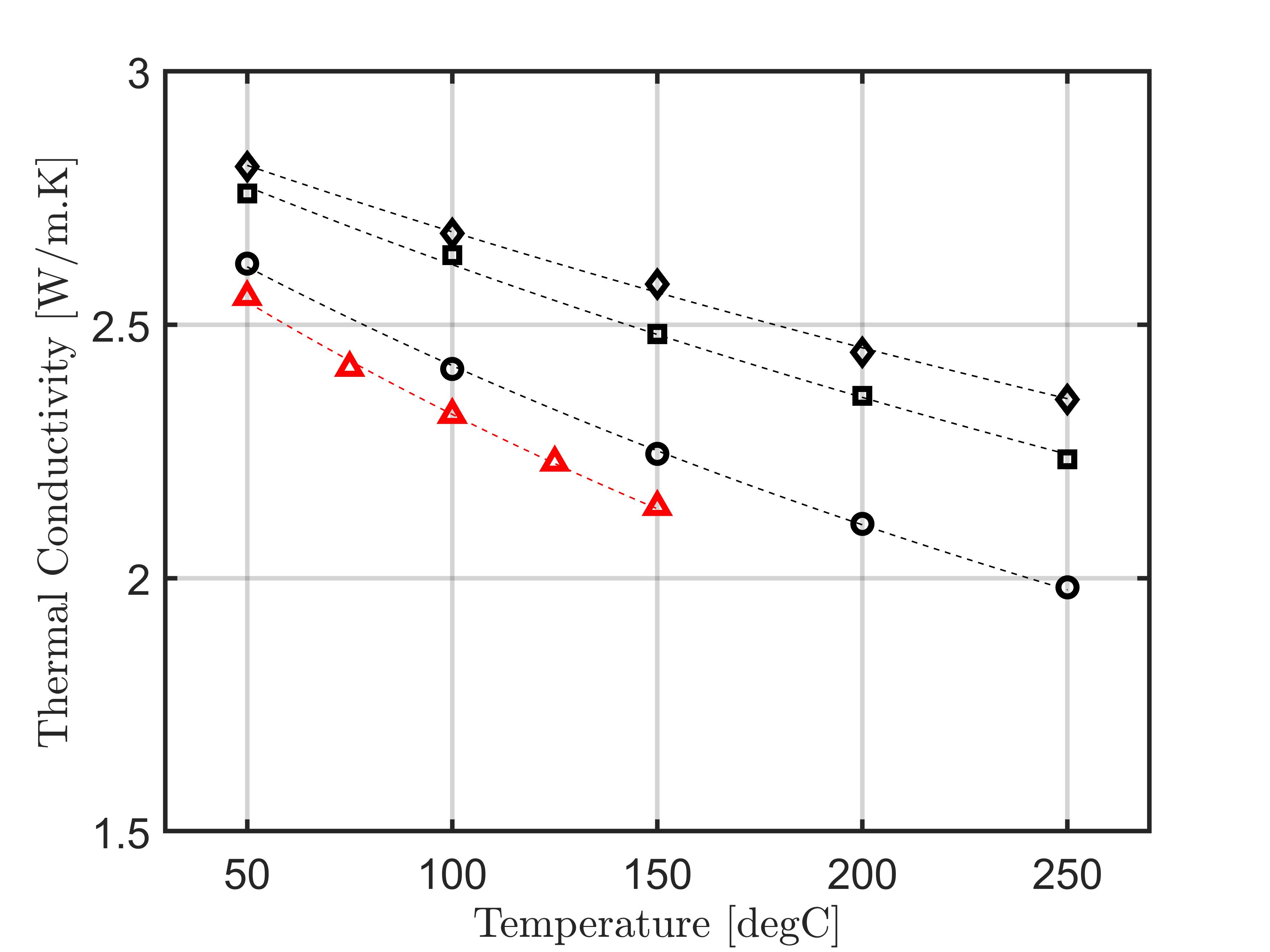}
        \subcaption{\label{fig:TC_S12}}
    \end{subfigure}
    \begin{subfigure}[t]{0.15\textwidth}
        \centering
        \includegraphics[width=\textwidth]{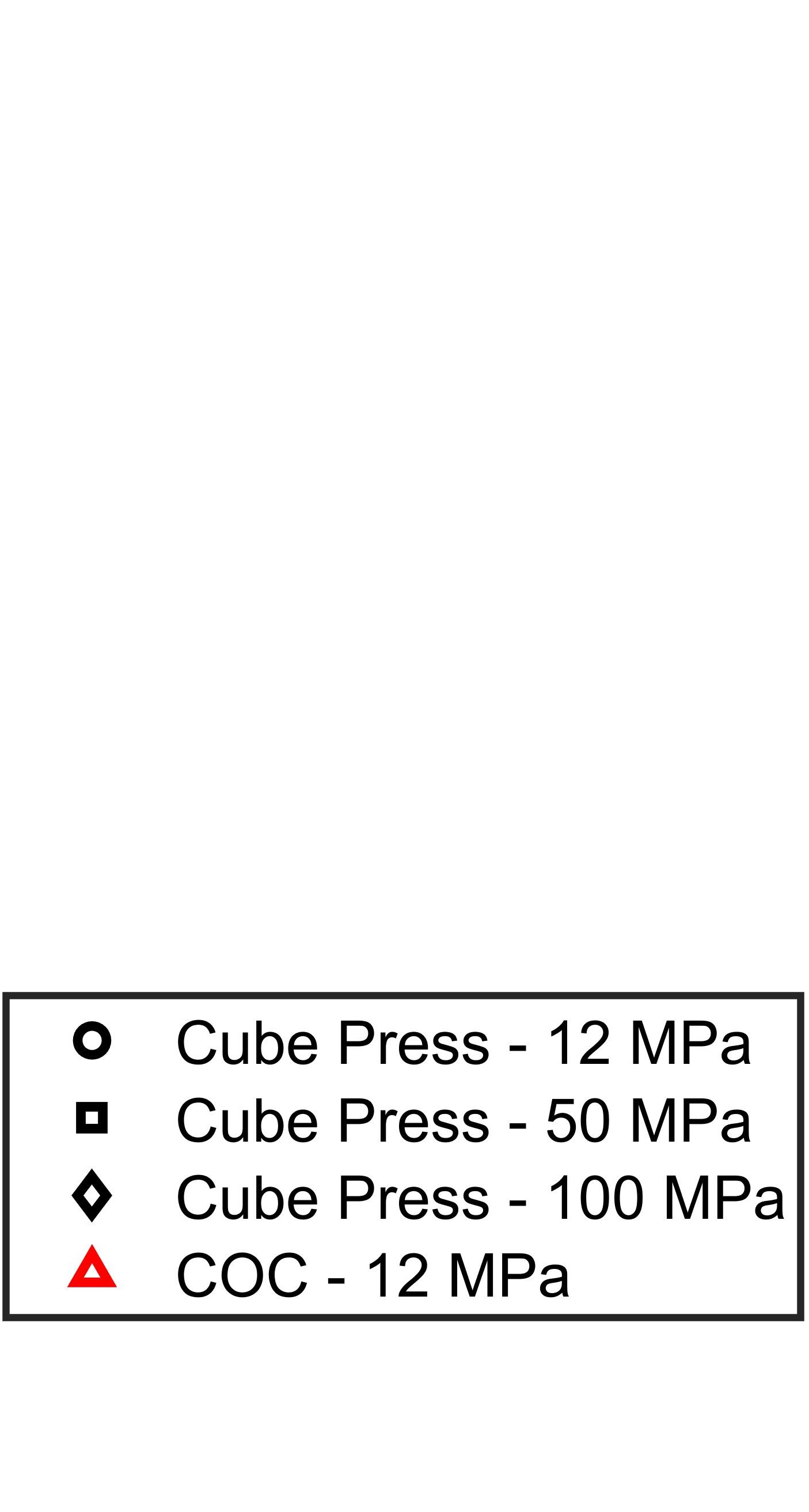}
    \end{subfigure}
    \caption{Thermal conductivity of samples S8 (left) and S12 (right) as a function of temperature, evaluated at different confining pressures. Results are shown for both the cube press configuration and the COC setup.}
    \label{fig:TC_s12_s8}
\end{figure}

Thermal conductivity for samples S\#8 (Fig.\ref{fig:TC_S8}) and S\#12 (Fig.\ref{fig:TC_S12}) is shown as a function of temperature for different loading geometries. Black symbols give cube-press measurements (3D loading) at 12, 50, and 100 MPa; red triangles show the COC configuration (uniaxial loading) plotted only at 12 MPa for direct comparison. As can be seen, the COC and cube-press results at 12 MPa are in good agreement across the entire temperature range, which further validates the cube-press implementation of the thermal conductivity measurement. The slight systematic offset,where cube-press values are mhigher than COC, could be attributable to the loading geometry: three-dimensional confinement promotes the intergranular contact, yielding higher effective conductivity than the uniaxial case. This trend is consistent for both S8 and S12. Furthermore, as expected for polycrystalline rocks, and in line with observations reported in the literature, thermal conductivity decreases with increasing temperature and increases with confining pressure. The reduction with temperature is attributed to enhanced phonon–phonon scattering and the gradual loss of intergranular contact efficiency as thermal expansion activates microcracks. The increase with pressure is explained by progressive closure of pores and cracks, which enhances solid–solid contact and improves heat transfer across grain boundaries. At higher pressures, once the most compliant cracks are closed, the incremental effect of pressure becomes less pronounced.

Before exploring the temperature effect in detail, it is worth noting that, to further increase confidence in the reported values, room-temperature measurements were performed using a commercial device (Thermtest TLS-100) based on the transient line heat-source method. The average thermal conductivity from five repeats was 2.11~$\text{W}~\text{m}^{-1}~\text{K}^{-1}$ for S\#8 and 2.5~$\text{W}~\text{m}^{-1}~\text{K}^{-1}$ for S\#12. These values are consistent with the absolute level of conductivity determined for the same materials in the experiments.

As outlined in the introduction, the thermal conductivity of a structurally perfect isotropic single crystal is theoretically expected to scale inversely with temperature, i.e. $k \propto T^{-1}$. However, according to \cite{emirov2021studies}, for multicomponent polycrystalline rocks with a large number of defects and dislocations, a weaker dependence of $k \propto \text{T}^{-0.5}$ is more appropriate. On the other hand, $k \propto \text{T}^{0.5}$ has been reported to describe the behavior of solids with an amorphous structure.

\begin{figure}[h!]
  \centering
  \begin{subfigure}[b]{0.4\textwidth}
    \includegraphics[width=\textwidth]{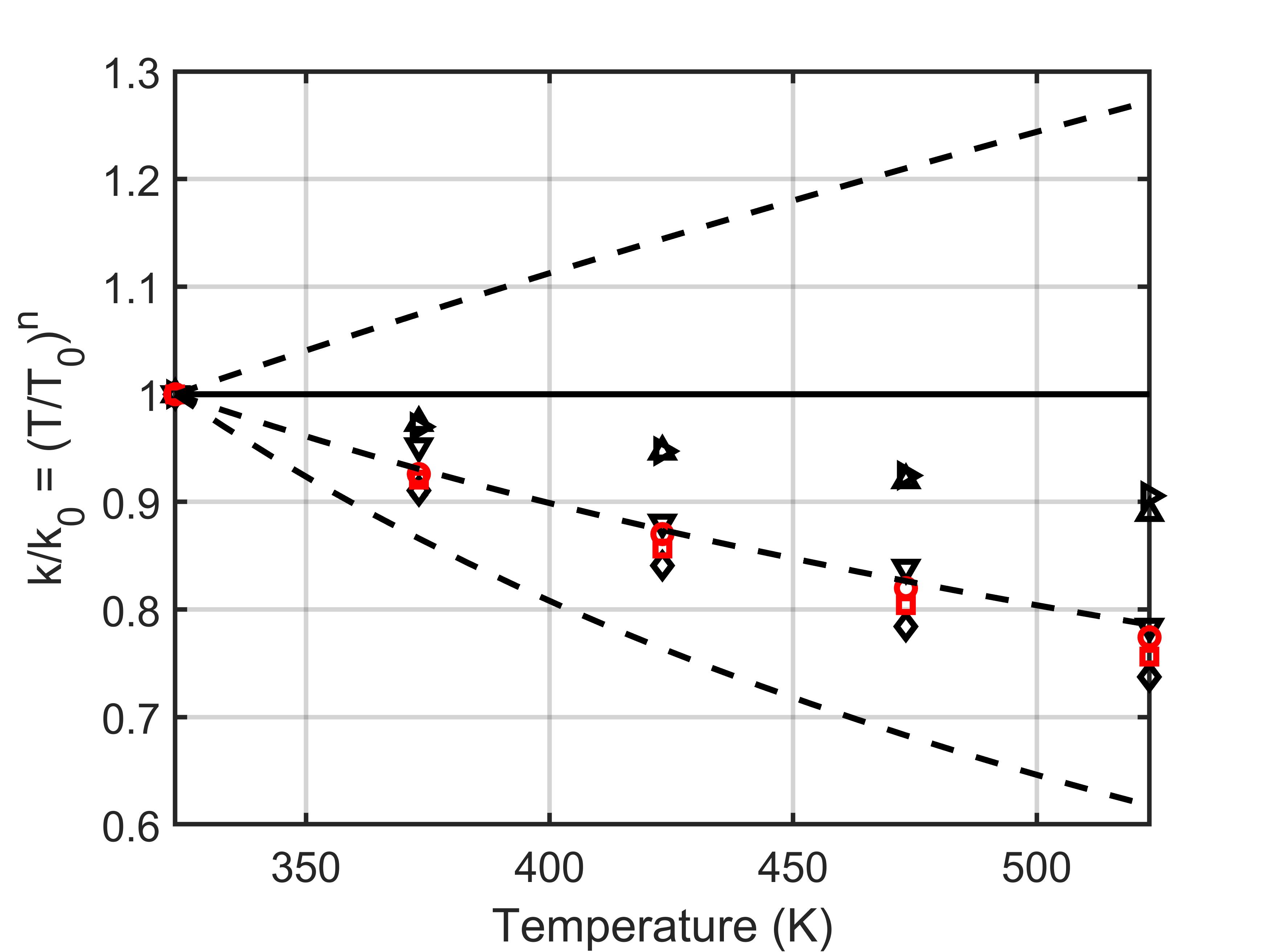}
    \subcaption{\label{fig:sandstone_literature_1}}
  \end{subfigure}
  \hfill
  \begin{subfigure}[b]{0.4\textwidth}
    \includegraphics[width=\textwidth]{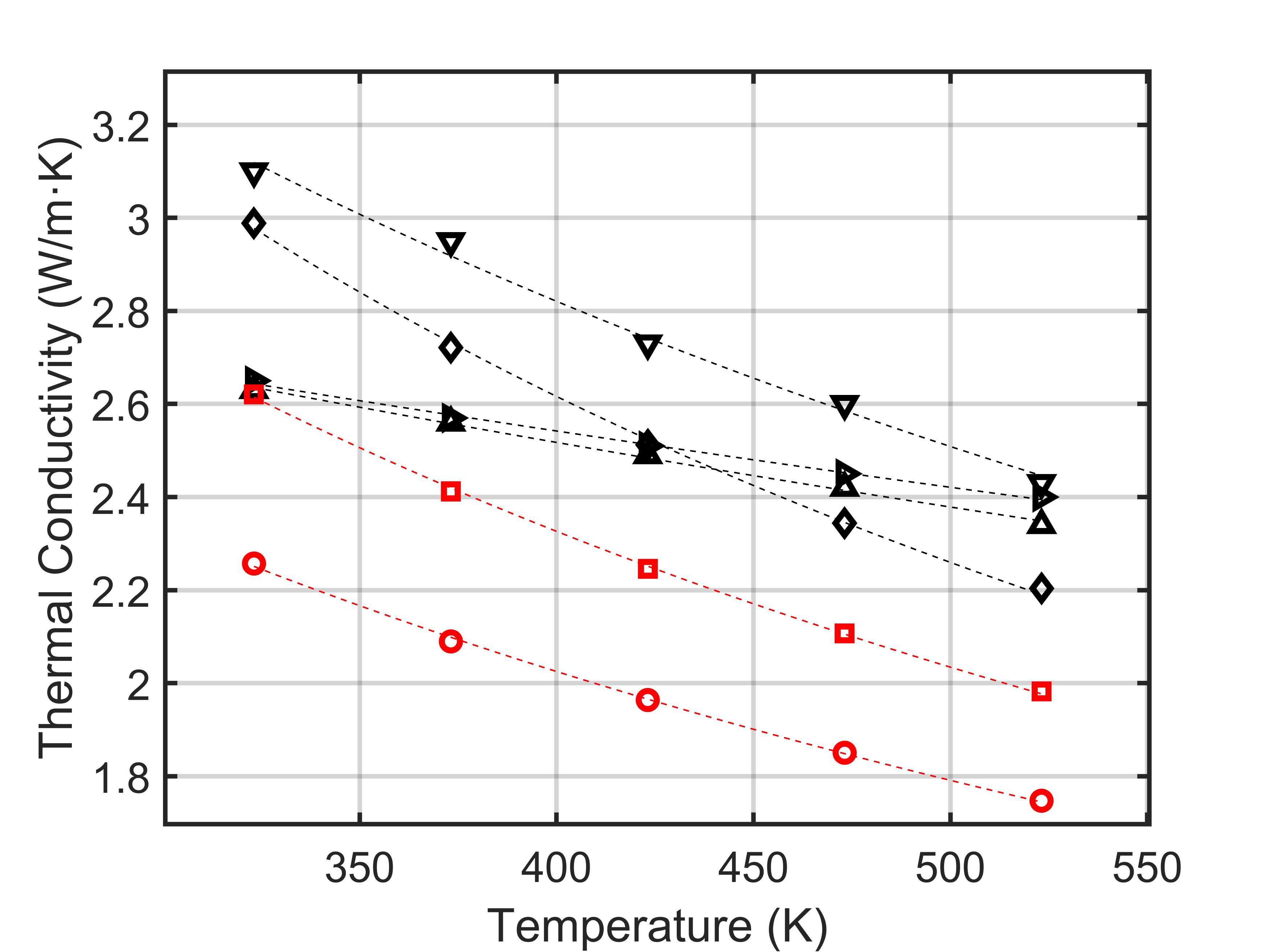}
    \subcaption{\label{fig:sandstone_literature_2}}
  \end{subfigure}
  \begin{subfigure}[t]{0.15\textwidth}
        \centering
        \includegraphics[width=\textwidth]{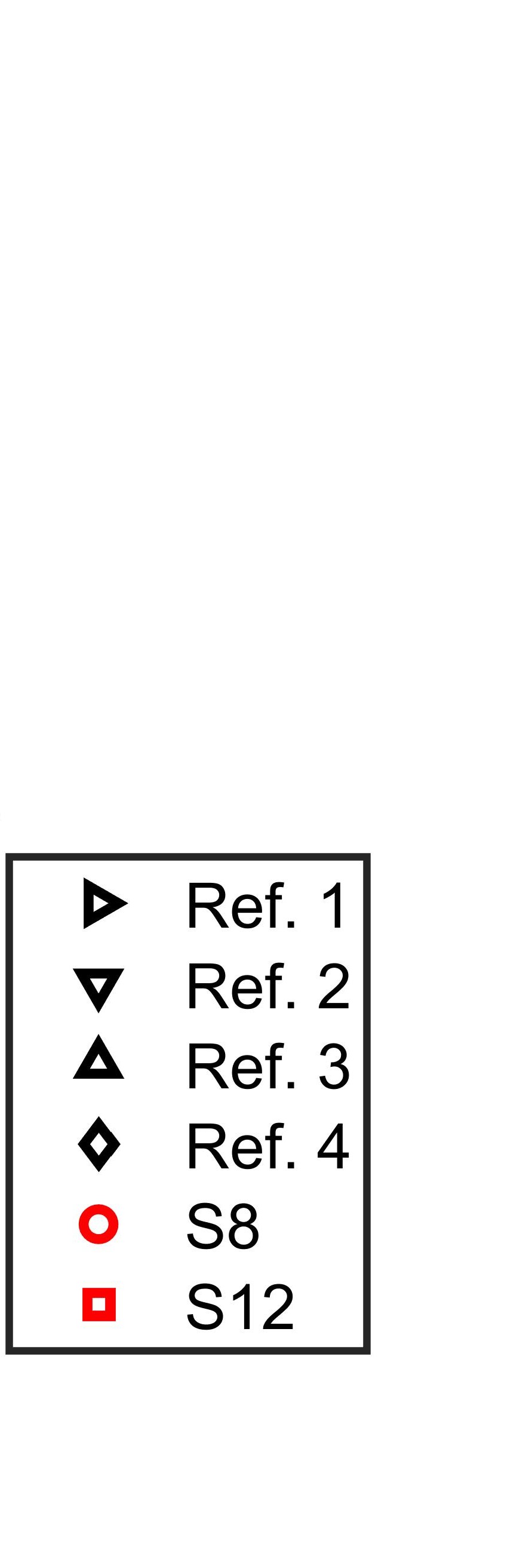}
    \end{subfigure}
  \caption{Comparison of the temperature-dependent thermal conductivity of sandstone samples evaluated in this study and gathered from literature. (a) Normalized values and (b) fitted using Eq.~\ref{eq:temp_dependence}; Ref~1: ~\cite{emirov2021studies}, Ref~2:~\cite{chen2021effect}, Ref~3:~\cite{abdulagatova2009effect}, and Ref~4:~\cite{miao2018temperature}.\label{fig:sandstone_literature}}

\end{figure}

To isolate the temperature dependence, conductivity values were normalized to those at $50~^{\circ}\text{C}$ under 12 MPa in the cube press. The normalized results are shown in Fig.~\ref{fig:sandstone_literature_1} together with a compilation of temperature-dependent thermal conductivity values of various sandstone samples gathered from the literature. These literature samples span a broad range of intrinsic properties, including differences in quartz content and porosity. For reference, additional distribution lines corresponding to $k \propto \text{T}^{-1}$, $k \propto \text{T}^{-0.5}$, and $k \propto \text{T}^{0.5}$ are also plotted. Additionally, the model introduced by \cite{kingery1959thermal} is one of the most widely used approaches for describing the temperature dependence of thermal resistivity in multi-component systems. It is expressed as:

\begin{equation}
\frac{1}{k(T)} = a + bT
\label{eq:temp_dependence}
\end{equation}

where $a$ and $b$ are empirical constants. According to \cite{vcermak1982thermal}, at temperatures below 600~\textdegree C, the thermal conductivity of rocks can be reliably described using Eq.~\ref{eq:temp_dependence}. Many researchers \citep{clauser1995thermal,seipold1995variation} have successfully fitted experimental data to this equation to capture the observed temperature dependence. Fig.~\ref{fig:sandstone_literature_2} presents a summary of the same gathered temperature-dependent thermal conductivity values of various sandstone samples, and the ones evaluated in this study, and evaluates the applicability of selected models, i.e., Eq.~\ref{eq:temp_dependence}, by fitting them to the observed trends. The comparison demonstrates that the observed temperature-dependent trends obtained in this study are consistent with those reported for sandstones in the literature.

\begin{figure}[h!]
  \centering
  \begin{subfigure}[b]{0.4\textwidth}
    \includegraphics[width=\textwidth]{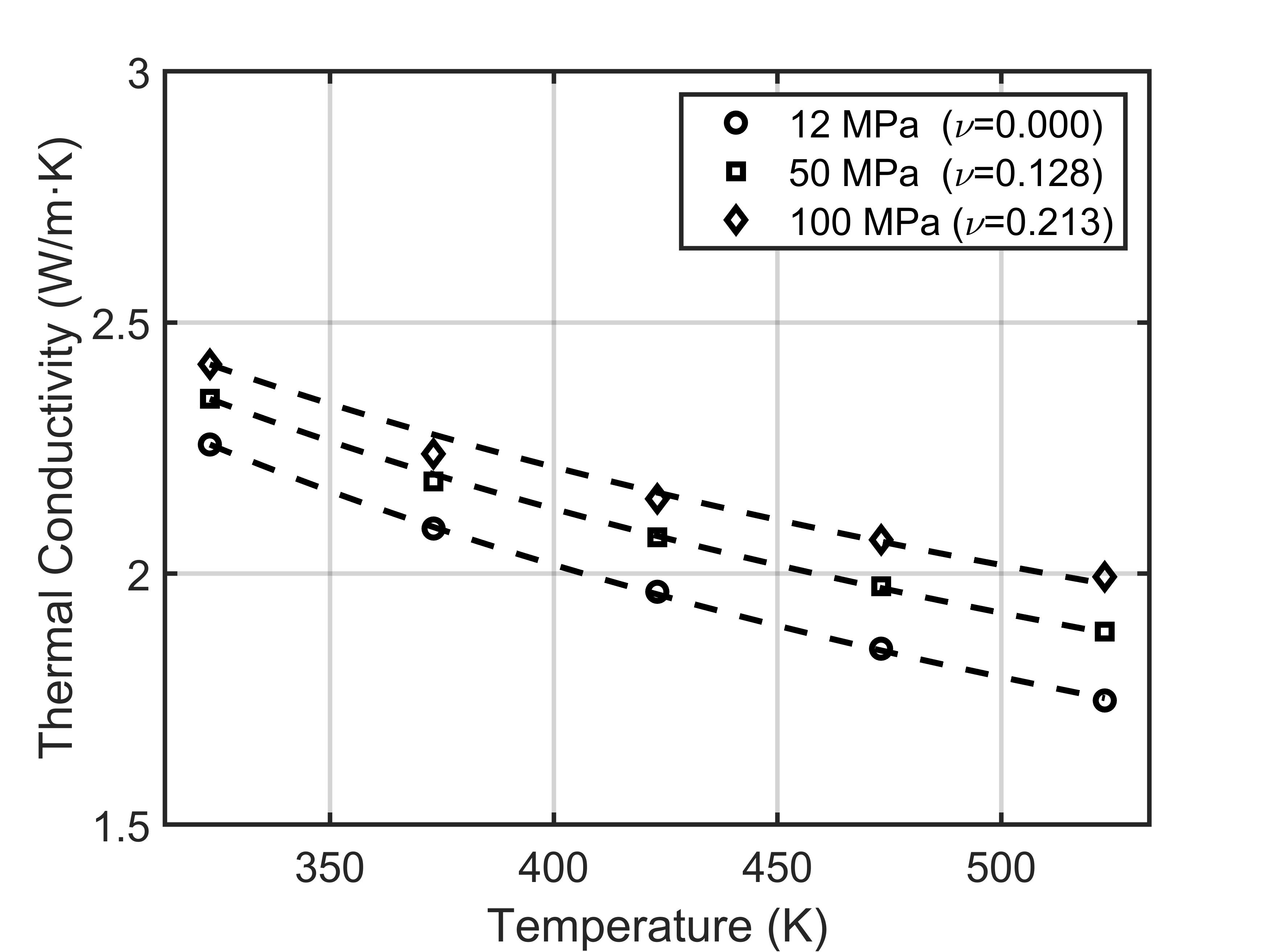}
    \subcaption{\label{fig:S8_TP}}
  \end{subfigure}
  \hfill
  \begin{subfigure}[b]{0.4\textwidth}
    \includegraphics[width=\textwidth]{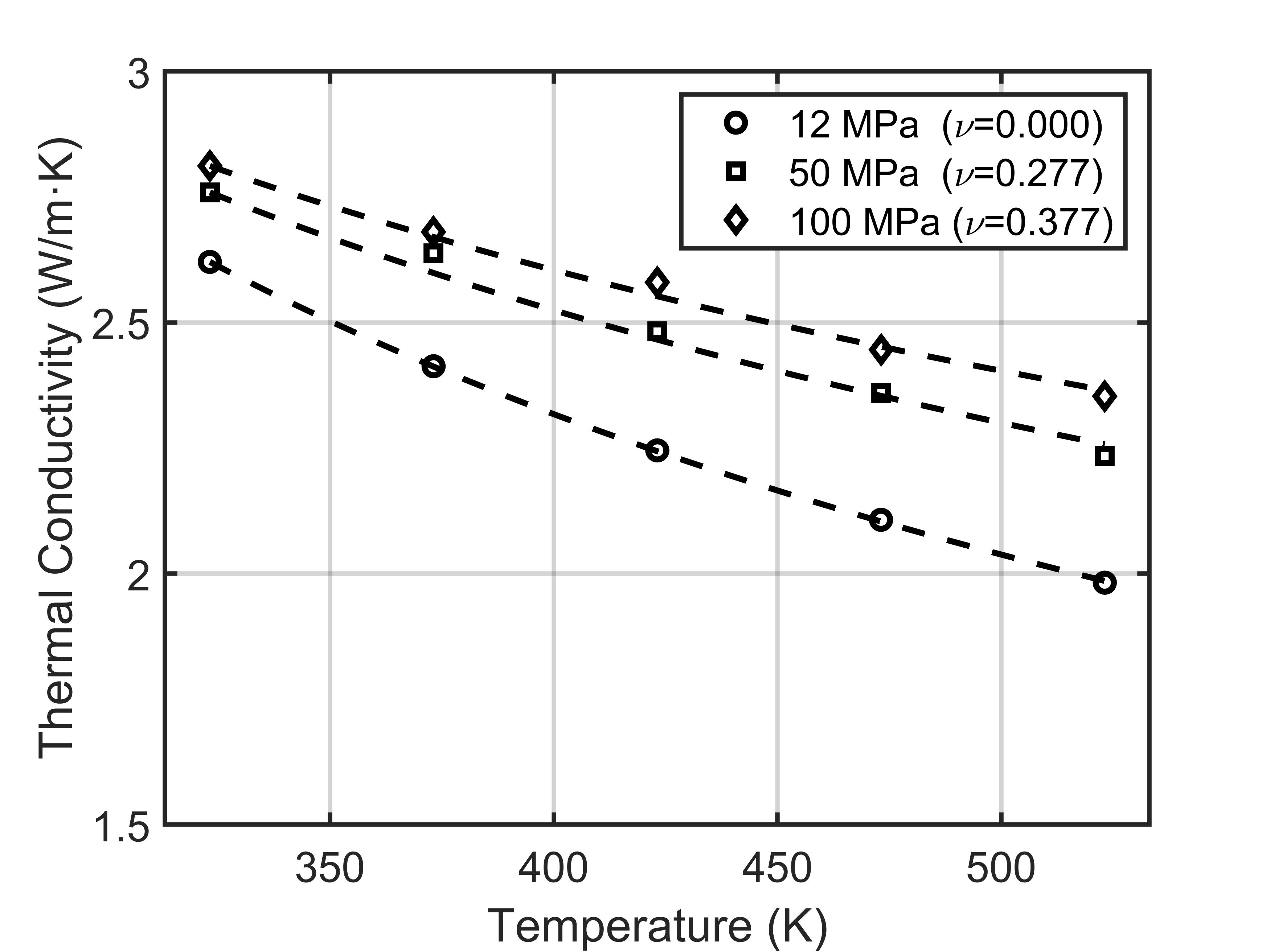}
    \subcaption{\label{fig:S12_TP}}
  \end{subfigure}
  \caption{Pressure–temperature dependence of thermal conductivity for sandstone samples S8 (left) and S12 (right) measured in the cube press. The data (symbols) are compared with the semi-empirical formulation (Eq.~\ref{eq:lambdaTP}), which accounts for the coupled effect of temperature and pressure. Dashed lines represent model fits.\label{fig:S8_S12_TP}}
  \end{figure}

Compared to temperature, the effect of pressure on thermal conductivity has received less attention in the literature. This is largely due to the common assumption that temperature exerts the most significant influence on thermal behavior. In higher pressure regimes ($P \ge 100~ \text{MPa}$), the pressure dependence of thermal conductivity in rocks can be reasonably approximated using linear models as 
\begin{equation}
    k(p) = k_0(1+\alpha p)
    \label{eq:pressure_effect}
\end{equation}
where $\alpha$ is a constant. While Eq.~\ref{eq:pressure_effect} has been shown to hold even at lower pressure ranges ($P < 100\text{MPa}$) for certain rock types~\citep{seipold1980measurements}, the pressure effect in this regime should not be neglected. At relatively low pressures, the closure of pores and microcracks enhances thermal contact between mineral grains, leading to a notable and often rapid increase in thermal conductivity. In this regard, \cite{horai1989effect} investigated the thermal conductivity of 23 silicate rocks under quasi-hydrostatic pressures up to 12~kbar. Their results indicate that thermal conductivity increases rapidly at pressures below 2~kbar, and follows an approximately linear trend at higher pressures for most samples. Furthermore, based on the data summarized by \cite{vcermak1982thermal} regarding the pressure dependence of sandstone thermal conductivity, an increase of approximately 4~\% per 100~MPa is observed between 4 and 200~MPa, while the increase averages around 0.5~\% per 100~MPa between 200 and 840~MPa under applied uniaxial pressure. Similarly, \cite{abdulagatova2009effect} reported that for sandstone, ETC increases by approximately 1.3–2.0~\% per 10~MPa at low pressures (below 100~MPa), whereas at higher pressures, the rate of increase diminishes to about 0.1–0.2~\% per 10~MPa. In this study, over the pressure range up to 100 MPa, an average increase of 1.14\% for S8 and 1.64\% for S12 across the entire investigated temperature interval was observed. As mentioned earlier, the effect of pressure is twofold. First, as already noted, an increase in confining pressure directly enhances thermal conductivity due to progressive crack closure and improved grain–grain contact. Second, pressure modifies the temperature dependence itself, such that the rate of conductivity decrease with increasing temperature becomes less pronounced at higher pressures. This coupled influence of pressure and temperature on thermal conductivity has not been comprehensively investigated in the literature, and the amount of experimental data available remains scarce. Consequently, a direct comparison across studies is limited. Therefore, in this work the results are compared to a semi-empirical formulation that has been introduced by \cite{emirov2021studies} to study the combined effects of pressure and temperature on rock thermal conductivity. Accordingly, the temperature-baric dependence of rock thermal conductivity can described as
\begin{equation}
    k(T,P) = k(T_0,P).(T/T_0)^{n_0.(1-\nu(p))}
    \label{eq:lambdaTP}
\end{equation}

where $\nu(p) = \frac{-1}{n_0}.\frac{\partial n}{\partial P}$ is the relative pressure coefficient of the exponent $n$. Here, $n_0$, defines the temperature dependence $k/k_0 = (T/T_0)^n$, at 12 Mpa of applied pressure. This relation has been shown to capture the gradual reduction in temperature sensitivity with increasing confining pressure. As illustrated in Fig.~\ref{fig:S8_S12_TP}, the measured data for both S8 and S12 can be well reproduced by the model. The results show that the coupled effect of pressure and temperature on thermal conductivity in these sandstones can be satisfactorily described within the framework of Eq.~\ref{eq:lambdaTP}, supporting its applicability to porous polycrystalline rocks.

\section{Conclusion}

Through the application of a comparative steady-state technique, this study presents a systematic dataset on the coupled influence of pressure and temperature on the thermal conductivity of low-porosity, dry sandstones. A unique aspect is the concurrent acquisition of three-dimensional P- and S-wave velocities, which constrain the sample’s microstructural evolution as it is subjected to varying pressure–temperature boundary conditions. To this end, a multianvil cubic press at the Geomechanics \& Geotechnics group, Kiel University, was adapted to impose in-situ–relevant conditions. Expected first-order trends were obtained: thermal conductivity decreased with temperature and increased with confining pressure, with both absolute values and normalized trends consistent with previously reported ranges. The obtained conductivity values were further validated through independent measurements using an in-house developed customized oedometer cell and a commercial device, confirming the reliability of the developed approach.

This initial implementation serves primarily as a methodological proof of concept. Although further optimization of temperature calibration and stability is anticipated to enhance precision, the demonstrated feasibility of simultaneous thermal-conductivity and wave-velocity measurements under coupled pressure and temperature conditions opens promising perspectives for studying more complex lithologies. The combined thermal–elastic approach provides a new experimental basis for linking heat transport to microstructural processes such as crack evolution and grain rearrangement, thereby supporting the calibration of thermo-mechanical models for crustal rocks.\newline
\break
\newpage
\bibliographystyle{elsarticle-harv} 
\bibliography{main.bib}
\end{document}